\documentclass{article}
\usepackage{amssymb}
\usepackage{bm}

\usepackage{graphicx}
\usepackage{amsmath}

\def\nsection#1{\section{#1}\setcounter{equation}{0}}

\def\nappendix#1{\vskip 1cm\noindent{\Large\bf Appendix
         #1}\def\thesection{#1}\setcounter{equation}{0}\vskip 0.3cm}

\newcommand{\qq}{\begin{eqnarray}}
\newcommand{\qqq}{\end{eqnarray}}
\newcommand{\ee}{{\rm e}}

\newcommand{\CV}{{\cal V}}

\newcommand{\CX}{{\cal X}}

\newcommand{\NR}{{\mathbf R}}

\begin{document}

\title{\textbf{Eulerian and Lagrangian pictures of non-equilibrium diffusions}}
\author{Rapha$\mathrm{\ddot{e}}$l Chetrite$^{1,2}$
\,\ and \,Krzysztof Gaw\c{e}dzki$^{1}$ \\ 
\\
\hspace*{-0.3cm}$^{1}$\small{Laboratoire de Physique, C.N.R.S., ENS-Lyon,
Universit\'e de Lyon, 46 All\'ee d'Italie,}\cr
\hspace*{-8.9cm}\small{69364 Lyon, France}
\hfill\cr
\hspace*{-0.3cm}$^{2}$\small {Physics of Complex Systems, 
Weizmann Institute of Science,
Rehovot 76100, Israel}\hfill}
\date{}
\maketitle

\vskip 0.3cm

\abstract{\noindent We show that a non-equilibrium diffusive dynamics 
in a finite-dimensional space takes in the Lagrangian frame of its mean local 
velocity an equilibrium form with the detailed balance property. This
explains the equilibrium nature of the fluctuation-dissipation relations
in that frame observed previously. The general considerations are 
illustrated on few examples of stochastic particle dynamics.}

\vskip 0.3cm
\

\nsection{Introduction}

In the last decades, non-equilibrium statistical mechanics has been 
a subject of intensive studies. One of the multiple aims 
of the research is the understanding of essential differences 
between the equilibrium and non-equilibrium dynamics. This is the
question that we shall address below. In the modelling of 
statistical-mechanical dynamics, an important role has been
played by stochastic Markov processes. Although largely idealized,
they often provide a sufficiently realistic description of experimental 
situations and have traditionally served as a playground for both 
theoretical considerations and numerical studies. The Markov processes 
corresponding to the equilibrium dynamics are characterized 
by the detailed balance property assuring that the net probability
fluxes between micro-states of the system vanish. On the other hand,
in the non-equilibrium Markov dynamics, the detailed balance is broken 
and there are non-zero probability fluxes even in a stationary situation.
\vskip 0.1cm
 
In the present paper, we shall consider only diffusive processes,
discarding Markov processes with discrete time or random jumps. 
For such systems, the detailed balance can be expressed as the vanishing
of the probability current that is non-zero in the non-equilibrium 
situations. It is convenient to represent the probability current in 
a hydrodynamical form as the instantaneous probability density of the 
process multiplied by the mean local velocity. The latter is the average 
instantaneous velocity of the process conditioned to pass through a 
given point. It will play the main role in what follows.
\vskip 0.1cm

In the past, there have been many attempts to apply ideas 
from statistical mechanics to the hydrodynamics of turbulent flows.
The success was limited by the fact that most methods of statistical physics 
had been developed for systems in or close to equilibrium whereas 
developed turbulence is a far-from-equilibrium phenomenon. Here we shall 
follow a reversed strategy, applying an idea from hydrodynamics to 
non-equilibrium statistical mechanics. There is a long tradition 
(going back to Lagrange) to describe the evolution of hydrodynamical 
fields in the Lagrangian frame that moves with fluid particles \cite{MY}. 
It is believed that such a description makes the intrinsic features of fluid 
dynamics at small scales more directly accessible than in the Eulerian 
(i.e. laboratory) frame. This is particularly true about the hydrodynamical 
advection that gains a simple representation in the Lagrangian frame. 
{\,\bf The main result of the present paper consists of a simple observation 
that the non-vanishing probability current in a Markov diffusion may 
be decoupled from the stochastic dynamics by passing to the Lagrangian 
frame of the mean local velocity.} \,More exactly, in the latter frame, 
the stochastic dynamics, although non-stationary, satisfies the detailed 
balance condition and the instantaneous 
probability density of the process does not change in time. The 
equilibrium-like Lagrangian-frame process does not contain information 
about the non-vanishing probability current of the original Eulerian-frame 
process but, if that information is provided independently, the 
Eulerian-frame process may be reconstructed from the Lagrangian-frame one. 
In short, the passage to the Lagrangian frame of the mean local velocity
re-expresses a non-equilibrium diffusion process as an equilibrium-type
one plus the decoupled probability current. To our knowledge, this 
rather straightforward observation about non-equilibrium diffusions has 
not been discussed in the literature, although a similar idea was recently 
employed in the quantum many-body dynamics \cite{Tokatly}.
\vskip 0.1cm

The paper consists of seven Sections and four Appendices. 
Sect.\,\ref{sec:Eulerpic} sets the stage and notations by briefly 
stating the basic definitions relevant for the diffusion processes that 
we consider. We introduce the notions of the probability current and 
of the mean local velocity and recall the concept of detailed balance. 
The crucial Sect.\,\ref{sec:Lagrpict} is devoted to the Lagrangian 
picture of diffusions. We define the Lagrangian frame of the mean 
local velocity and compute the instantaneous probability density of 
the Lagrangian-frame process. By working out the stochastic differential 
equation satisfied by this process, we show that it is a non-stationary 
diffusion with the detailed balance property. Two simple examples
illustrate the general considerations: a diffusion of a particle on 
a circle in the presence of a constant force and a linear stochastic 
equation describing a Rouse model of a polymer in shear flows. We also
discuss the reconstruction of the original Eulerian-frame process from 
the Lagrangian-frame one. Sect.\,\ref{sec:Langev} is devoted to the 
Langevin equations with both Hamiltonian and non-conservative forces. 
In this case, it is convenient to modify the definition of the probability 
current and the mean local velocity to assure that they vanish in the 
absence of the non-conservative drift. The main properties of the 
Lagrangian-frame process are unaffected by this modification. We 
illustrate the general discussion by the example of a harmonic chain. 
Sect.\,\ref{FDT} discusses the extensions of the Fluctuation-Dissipation 
Theorem to the non-equilibrium situation in the light of the results
about the Lagrangian-frame process. These results provide a deeper reason 
for the observation made in \cite{CFG}, see also \cite{SpS0}, that
the fluctuation-dissipation relations takes the equilibrium form
in the Lagrangian frame of the mean local velocity. 
In Sect.\,\ref{sec:obstr}, we point out that important non-equilibrium 
diffusion processes in infinite-dimensional spaces, like the one-dimensional 
KPZ equation or the processes describing the large-deviations regime of
fluctuations around the hydrodynamical limit of the boundary-driven 
zero-range particle processes do not possess Lagrangian picture. Finally, 
Sect.\,\ref{sec:concl} presents our conclusions. Appendices collect some 
more technical arguments.
\vskip 0.2cm

\noindent{\bf Acknowledgements}. \ The authors thank Gregory Falkovich
for discussions. R.C. acknowledge the support of the Koshland Center 
for Basic Research and K.G. of the project ANR-05-BLAN-0029-03.

\nsection{Eulerian picture of diffusions}
\label{sec:Eulerpic}
\subsection{Diffusion processes}

\noindent We shall begin by considering a general diffusion process 
$\,x_t\,$ in a $\,d$-dimensional (phase-)space $\,\CX\,$ with coordinates 
$\,(x^i)$, \,of the same type as in ref. \cite{CG1} that was 
devoted to the study of fluctuation relations for such processes.
The examples we shall have in mind include various types of Langevin 
dynamics used to model equilibrium and non-equilibrium dynamics as well 
as the Kraichnan model of turbulent advection \cite{Warw}. Of the
rich theory of diffusion processes, see e.g. \cite{Risk,Oksen,Stroock},
we shall need only few basic facts that we collect below. The process 
$\,x_t\,$ is assumed to satisfy the stochastic differential equation (SDE)
\qq
\dot{x}_t\ =\ u_t(x_t)\,+\,\zeta_t(x_t)\,,  
\label{SDE}
\qqq
where $\,\dot{x}_t\equiv\frac{dx_t}{dt}$ \,and, on the right hand side, 
$\,u_t(x)\,$ is a time-dependent deterministic vector field (the drift), 
and $\,\zeta_t(x)\,$ is a Gaussian random vector field with mean zero 
and covariance
\qq
\big\langle\, \zeta^{i}_t(x)\ \zeta^{j}_{s}(y)\,\big\rangle\ =\ 2
\,\delta(t-s)\,D_{t}^{ij}(x,y)\,.  \label{dep2}
\qqq
Note that $\,\zeta_t(x)\,$ is a white noise in time so that
Eq.\,(\ref{dep2}) requires a choice of 
a stochastic convention. As in \cite{CG1}, we shall interpret it
in the Stratonovich sense to assure that $\,u^i_t(x)\,$ and 
$\,\zeta^i_t(x)\,$ transform as vector fields under a change of 
coordinates\footnote{\parbox[t]{15.2cm}{In probabilists'
notations, Eq.\,(\ref{SDE}) would read $\ dx_t=u_t(x_t)\,dt
+\sum\limits_nX_n(x_t)\circ dW^n_t\ $ where $\,X_n\,$ are vector fields 
such that $\,2D^{ij}(x,y)=\sum\limits_nX^i_n(x)X^j_n(y)\,$ 
and $\,W^n_t\,$ are independent Wiener processes.}}.
\,The single time expectations of functions of the process
$\,x_t\,$ evolve according to the equation
\qq
\frac{d}{{dt}}\,\big\langle\,f(x_t)\,
\big\rangle\,=\,\big\langle\,(L_tf)(x_t)\,\big\rangle\,,\ \qquad{\rm where}
\,\ \qquad
L_t\,=\,\hat{u}^{i}_t\partial _{i}+\partial
_{j}d_{t}^{ij}\partial _{i}
\label{Generator}
\qqq
\vskip -0.3cm
\noindent with 
\vskip -0.4cm
\qq
d_{t}^{ij}(x)\,=\,D_{t}^{ij}(x,x)\qquad  
\hat{u}^i_t(x)\,=\,u^i_t(x)\,-\,r^i_t(x)\,,\qquad
r^i_t(x)\,=\,\partial_{y^j}D_t^{ij}(x,y)|_{y=x}
\label{stand}
\qqq
are the instantaneous generators of the process $\,x_t$. \,Note the presence
of the term $\,r_t\,$ correcting the drift and due to the dependence 
the covariance of $\,\zeta_t\,$ on the points in $\,\CX$. 
\,The time evolution of the instantaneous (i.e. single-time) probability 
density function (PDF) of the process
\qq
\rho_t(x)\ =\ \big\langle\,\delta(x-x_t)\,\big\rangle
\label{instden}
\qqq
is governed by the formal adjoints $\,L_t^\dagger\,$ of the
generators $\,L_t\,$: 
\qq
\partial_t\rho_t\ =\ L_t^\dagger\rho_t\ =\ 
-\partial_i\big[\hat u^i_t\rho_t\,-\,d^{ij}_t
\,\partial_j\rho_t\big]\,.
\label{evrho}
\qqq The transition PDF's of the Markov process $\,x_t\,$ given by
the conditional expectations
\qq
P(s,x;t,y)\ =\ \rho_s(x)^{-1}\,
\big\langle\,\delta(x-x_s)\,\delta(y-x_t)\,\big\rangle
\qqq
with $\,s\leq t\,$ satisfy the Chapman-Kolmogorov composition
rule $\ \int P(r,x;s,y)\,P(s,y;t,z)\,dy\,=\,P(r,x;t,z)\ $
and the Kolmogorov differential equations
\qq
\partial_sP(s,x;t,y)\ =\ -L_s(x)\,P(s,x;t,y)\,,\qquad\partial_tP(s,x;t,y)
\ =\ L^\dagger_t(y)\,P(s,x;t,y)\,.
\label{Kolmo}
\qqq
The latter, together with the condition $\,P(t,x;t,y)=\delta(x-y)$,
\,determine the transition probabilities under appropriate regularity
assumptions \cite{Stroock}. 
\vskip 0.2cm

\subsection{Probability current and mean local velocity}

\noindent Some other basic notions concerning Markov
diffusions will play a central role below. The evolution 
equation (\ref{evrho}) for the instantaneous PDF (\ref{instden}) 
of the process $\,x_t\,$ has the form of the continuity 
equation
\qq
\partial_t\rho_t\ +\ \nabla\cdot j_t\ =\ 0
\label{conteq}
\qqq
with the {\,\bf probability current}
\qq
j^i_t\ =\ \big[\hat u^i_t\,-\,d^{ij}_t\,\partial_j\big]\rho_t
\label{curr}
\qqq
whose flux through the boundary of any region $\,\CV\,$ gives the rate
of change of the probability that $\,x_t\,$ belongs to $\,\CV$.
\ A more transparent interpretation of the current $\,j_t(x)\,$
is given by the formula: 
\qq
j^i_t(x)\ =\ \lim\limits_{\epsilon\to0}\, 
\,\Big\langle\,\frac{{x^i_{t+\epsilon}-x^i_{t-\epsilon}}}{{2\epsilon}}\,\,
\delta(x-x_t)\,\Big\rangle\ \equiv\ \big\langle\,\dot{x}^i_t\,\,\delta(x-x_t)
\,\big\rangle\,.
\label{prdel}
\qqq
that is proven in Appendix A. \,For it to hold, it is essential 
to use the symmetric derivative over time of $\,x_t\,$ because
the left and right time derivatives lead to different results,
with the difference coming from the white noise contribution to 
$\,\dot{x}_t\,$ \cite{Nelson}.
\vskip 0.1cm

The probability current $\,j^i_t(x)\,$ may be written in the form
borrowed from hydrodynamics as $\,\rho_t(x)\,v^i_t(x)\,$ where
\qq
v^i_t(x)\ =\ \rho_t(x)^{-1}j^i_t(x)\ =\ \frac{\big\langle\,\dot{x}_t\,\,
\delta(x-x_t)\,\big\rangle}{\big\langle\,\delta(x-x_t)\,\big\rangle}\ =\ 
\hat u^i_t(x)\,-\,d^{ij}_t(x)\,\partial_j\ln{\rho_t(x)}
\label{mlv}
\qqq
has the interpretation of the time $\,t\,$ mean velocity of the process 
conditioned to be at point $\,x\,$ (once again, the velocity 
should be defined by the symmetric time derivative). \,Accordingly,
the quantity $\,v_t(x)\,$ is called the {\,\bf mean local velocity}.
\,Geometrically, $\,v_t\,$ is a time dependent vector field on $\,\CX$,
\,as we show in Appendix B.
\,The continuity equation (\ref{conteq}) takes now a hydrodynamical form 
of the advection equation
\qq
\partial_t\rho_t\,+\,\nabla\cdot\big(\rho_tv_t\big)\ =\ 0
\label{adveq}
\qqq 
for the density $\,\rho_t(x)\,$ transported by the velocity field $\,v_t(x)$.
\vskip 0.1cm

The vanishing of the probability current $\,j_t(x)\,$ for densities
$\,\rho_t$,  \,or of the related mean local velocity $\,v_t(x)$, 
is usually taken as the definition of the {\,\bf detailed balance\,} 
for the process $\,x_t$. \,It assures that the instantaneous PDF 
of $\,x_t\,$ is time-independent: $\,\rho_t\equiv\rho$.
\,Assuming the detailed balance and introducing the Hamiltonian 
$\,H(x)=-\beta^{-1}\ln{\rho(x)}+const.$,
\,where $\,\beta^{-1}\,$ is the temperature in the energy units,
\,the SDE (\ref{SDE}) may be rewritten as the equilibrium-type Langevin 
equation 
\qq
\dot{x}_t^i\ =\ -\beta\,d^{ij}_t(x_t)
\,(\partial_jH)(x_t)\ +\ r^i_t(x)\ +\ \zeta^i_t(x)\,,
\label{equileq}
\qqq
with the notations of (\ref{stand}). Conversely, a dynamics
governed by equation (\ref{equileq}) satisfies the detailed balance
relative to the Gibbs density $\,Z^{-1}\ee^{-\beta\,H(x)}$,
\,where $\,Z\,$ (the partition function) is the normalization factor.  
Thus the equilibrium form (\ref{equileq}) of the dynamics is equivalent 
to the vanishing of the mean local velocity, the property independent 
of the choice of coordinate system. The presence of the correction 
$\,r_t\,$ in Eq.\,(\ref{equileq}) assures that the drift term transforms
as a vector field under a change of coordinates if $\,\ee^{-\beta H}\,$ 
transforms as a density, \,see Eq.\,(\ref{B10}) in Appendix B.
\vskip 0.1cm

The above general considerations carry over, at least on an informal level, 
to diffusion processes in infinite-dimensional spaces described by stochastic 
partial differential equations. Nevertheless, as explained 
in Sect.\,\ref{sec:obstr}, in few important examples of infinite-dimensional 
non-equilibrium diffusions there are obstructions to the realization 
of the part of our program that we discuss in the next section.

\nsection{Lagrangian picture of diffusions}
\label{sec:Lagrpict}

\subsection{Lagrangian frame of mean local velocity}
\label{sec:Lagrfr}

Recall that in hydrodynamics the motion of fluid particles in 
the Eulerian velocity field $\,v_t(x)\,$ is described by the ordinary 
differential equation
\qq
\dot{x}\ =\ v_t(x)\,
\label{lagrtr}
\qqq
that generates the flow $\,x\,\mapsto\,\Phi_t(x)\,$ assigning
to the initial condition $\,x\,$ of the fluid particle at time $\,t_0\,$
its position at time $\,t$. \,One has:
\qq
\partial_t\Phi_t(x)\ =\ v_t(\Phi_t(x))\qquad{\rm and}\qquad
\Phi_{t_0}(x)\ =\ x\,.
\label{Phieq}
\qqq   
We assume below that $\,\Phi_t\,$ is well defined for all times, 
see, however, Sect.\,\ref{sec:obstr}. \,The passage to the Lagrangian 
frame of the velocity field $\,v_t\,$ is realized by the family of inverse 
transformations $\ x\mapsto\Phi_t^{-1}(x)\ $ retracing back 
the flow. We have assumed that the Lagrangian
and the Eulerian frames coincide at time $\,t_0$. 
\vskip 0.1cm

Let us apply the above hydrodynamical idea to the diffusion
process $\,x_t$, \,describing it in the Lagrangian frame of the mean 
local velocity $\,v_t(x)$.  \,In this frame, the process $\,x_t\,$ 
becomes  
\qq
\tilde x_t\ =\ \Phi_t^{-1}(x_t)\,.
\qqq
In words, $\,\tilde x_t\,$ is the point that the particle of the
hypothetical fluid moving with the mean local velocity occupied
at time $\,t_0\,$ if at time $\,t\,$ it is at $ \,x_t$.
\,We shall show that the Lagrangian-frame stochastic process 
$\,\tilde x_t\,$ is again a diffusion by finding the SDE that it obeys.

\subsection{Instantaneous densities in the Lagrangian picture}
\label{sec:instdens}

\noindent Let us start by addressing the question what are
the instantaneous PDF's of the Lagrangian-frame process $\,\tilde x_t$. 
\,These are defined as
\qq
\tilde\rho_t(\tilde x)\ =\ \big\langle\,\delta(\tilde x-\tilde x_t)\,
\big\rangle\ =\ \big\langle\,\delta(\tilde x-\Phi_t^{-1}(x_t))\,
\big\rangle\,.
\qqq
Changing variables inside the delta-function on the right hand side, 
we may rewrite the above relation as the identity 
\qq
\tilde\rho_t(\tilde x)\ =\ \varphi_t(\tilde x)\,
\big\langle\,\delta(\Phi_t(\tilde x)-x_t)\,
\big\rangle\ =\ \varphi_t(\tilde x)\,\rho_t(\Phi_t(\tilde x))\,,
\label{cteden}
\qqq
where $\,\varphi_t\,$ is the Jacobian of the transformation $\,\Phi_t\,$:
\qq
\varphi_t(\tilde x)\ =\ \det\big((\partial_j\Phi_t)^i(\tilde x)\big)\ =\ 
\Big(\det\big(\partial_i\Phi^{-1}_t)^j\big)\big(\Phi_t(\tilde x)\big)
\Big)^{-1}\,.
\label{jac}
\qqq 
On the other hand, it is well known (and easy to check) that the solution 
of the Cauchy problem for the advection equation (\ref{adveq}) may be
written in the form
\qq
\rho_t(x)\ =\ \int\delta(x-\Phi_t(y))\,\rho_{t_0}(y)\,dy\ =\ 
\varphi_t(\tilde x)^{-1}\,\rho_{t_0}(\tilde x)\,.
\label{eq2}
\qqq
for $\,\tilde x=\Phi_t^{-1}(x)$. \,In words, Eq.\,(\ref{eq2}) states that 
$\,\rho_t(x)\,$ is equal to the
density $\,\rho_{t_0}(\tilde x)\,$ at the initial point of the Lagrangian
trajectory passing through $\,x\,$ at time $\,t$, \,divided by the factor 
$\,\varphi_t(\tilde x)\,$ giving the volume contraction around that
trajectory. \,Comparing Eqs.\,(\ref{cteden}) and
(\ref{eq2}), we infer that 
\qq
\tilde\rho_t(\tilde x)\ =\ \rho_{t_0}(\tilde x)\,.
\qqq
This shows that {\,\bf the instantaneous PDF's freeze in the Lagrangian frame 
to the time} $\,t_0\,$ {\bf value of the Eulerian-frame density}. 
\,Since the process $\,\tilde x_t\,$ itself is, in general, non-stationary, 
this might come as a surprise, although it is a direct consequence of 
the advection equation (\ref{adveq}).

\subsection{Stochastic equation for the Lagrangian-frame process}
\label{sec:Lfstocheq}

\noindent There are further surprises in the Lagrangian frame 
resulting in a simplification of the non-equilibrium dynamics.
Let us find the stochastic equation obeyed by the process $\,\tilde x_t$.
\,This is a straightforward, although somewhat tedious, exercise.
By the standard chain rule, that holds for the Stratonovich stochastic
equations,
\qq
\dot{\tilde x}_t^i\ =\ (\partial_t\Phi^{-1}_t)^i(x_t)\ 
+\ (\partial_k\Phi^{-1}_t)^i(x_t)\,\dot{x}^k_t\,.
\label{tosubs}
\qqq
Differentiating over time the identity 
$\ \Phi^{-1}_t\big(\Phi_t(\tilde x)\big)=\tilde x\ $
and setting $\ x=\Phi_t(\tilde x)$, \ we infer the relation
\qq
(\partial_t\Phi^{-1}_t)^i(x)\ =\ -(\partial_k\Phi^{-1}_t)^i(x)\,v^k_t(x)
\ =\ -(\partial_k\Phi^{-1}_t)^i(x)\,\big[
\hat u^k_t(x)\,
-\,d^{kl}_t(x)\,\partial_l\ln{\rho_t(x)}\big]
\,.
\label{explfor}
\qqq
The substitution of the last equality and of Eq.\,(\ref{SDE}) to the identity 
(\ref{tosubs}) gives:
\qq
\dot{\tilde x}_t^i&=&(\partial_k\Phi^{-1}_t)^i(x_t)\Big[-\hat u^k_t(x_t)\,
+\,d^{kl}_t(x_t)\,\partial_l\ln{\rho_t(x_t)}\,+\,u^k_t(x_t)
\,+\,\zeta^k_t(x_t)\Big]\cr\cr
&=&(\partial_k\Phi^{-1}_t)^i(x_t)\Big[r^i_t(x_t)\,
+\,d^{kl}_t(x_t)\,\partial_l\ln{\rho_t(x_t)}\,+\,\zeta^k_t(x_t)\Big],
\label{coss}
\qqq
where the second equality follows from Eqs.\,(\ref{stand}). \,Note the 
disappearance of the drift $\,u_t\,$ from the right hand side. 
\,Let us introduce the Lagrangian-frame white-noise vector field
\qq
\tilde\zeta^i_t(\tilde x)\ =\ (\partial_k\Phi^{-1}_t)^i(x)\,\zeta^k_t(x)
\qqq
for $\,x=\Phi_t(\tilde x)$. \,It has mean zero and covariance
\qq
\big\langle\,\tilde\zeta^i_t(\tilde x)\ \tilde\zeta^j_s(\tilde y)
\,\big\rangle\ =\ 2\,\delta(t-s)\,\tilde D^{ij}_t(\tilde x,\tilde y)
\qqq
\vskip -0.25cm
\noindent with
\vskip -0.45cm
\qq
\tilde D^{ij}_t(\tilde x,\tilde y)\ =\ (\partial_k\Phi^{-1}_t)^i(x)\,
D^{kl}_t(x,y)\,(\partial_l\Phi^{-1}_t)^j(y)
\qqq
for $\,x=\Phi_t(\tilde x)\,$ and $\,y=\Phi_t(\tilde y)$. \,Observe that
the covariances $\,\tilde D^{ij}_t\,$ and $\,D^{ij}_t\,$ are related
by the standard tensorial rule of transformation under the map 
$\,\Phi_t^{-1}$. \,We shall need two identities that may be obtained from 
the change-of-variables relations (\ref{a:r}) and (\ref{a:ln}) of Appendix B
if we set $\,\Psi=\Phi_t^{-1}\,$ there. \,They are:
\qq
\tilde r'^i_t(\tilde x)\ \equiv\ \partial_{\tilde y^j}D'^{ij}_{t}(\tilde x,
\tilde y)|_{\tilde y=\tilde x}\ =\ 
(\partial_k\Phi_t^{-1})^i(x)\,\Big[r^k_t(x)\,+\,
(\partial_{j}\Phi_t)^h(\tilde x)\,d^{kl}_t(x)\,
(\partial_h\partial_l\Phi_t^{-1})^j(x)\Big]\quad\label{za:r}
\qqq
\vskip -0.2cm
\noindent and
\vskip -0.55cm
\qq
(\partial_l\Phi_t^{-1})^j(x)\,(\partial_{j}\ln{\tilde\rho_t})(\tilde x)
\ =\ (\partial_l\ln{\rho_t})(x)\ -\ 
\,(\partial_{j}\Phi_t)^h(\tilde x)\,(\partial_l\partial_h\Phi_t^{-1})^j(x)\,.
\quad
\label{za:ln}
\qqq
Adding the first of the latter equations to the second one 
multiplied by $\ (\partial_k\Phi_t^{-1})^i(x)\,d^{kl}_t(x)$, \ we obtain 
the identity 
\qq
\tilde r'^i_t(\tilde x)\,+\,\tilde d^{ij}_t(\tilde x)\,
(\partial_{j}\ln{\tilde\rho_t)(\tilde x)}\ =\ 
(\partial_k\Phi_t^{-1})^i(x)\,\Big[r^k_t(x)\,+\,d^{kl}_t(x)\,
(\partial_l\ln{\rho_t})(x)\Big].
\label{alfin}
\qqq
Recalling that $\,\tilde\rho_t\equiv\rho_{t_0}\,$ for all $\,t\,$ and defining
the Lagrangian-frame Hamiltonian by the relation
\qq
\tilde H(\tilde x)\ =\ -\beta^{-1}\ln{\rho_{t_0}(\tilde x)}\,+\,const.
\qqq
for an arbitrary constant, \,the identity (\ref{alfin}) permits to rewrite 
the stochastic equation (\ref{coss}) in the form of Eq.\,(\ref{equileq}): 
\qq
\dot{\tilde x}^i_t\ =\ -\beta\,\tilde d^{ij}_t(\tilde x_t)\,(\partial_j\tilde H)
(\tilde x_t)\,+\,\tilde r^i_t(\tilde x_t)\,+\,\tilde\zeta^i_t(\tilde x_t)\,.
\label{tildequileq}
\qqq
This is the main result of this section: \,{\bf the Lagrangian 
frame process} $\,\tilde x_t\,$ {\bf satisfies the equilibrium Langevin 
equation with detailed balance relative to the density} 
$\,\rho_{t_0}(\tilde x)\ =\ Z^{-1}\ee^{-\beta\,\tilde H(\tilde x)}\,$ that stays 
invariant in the Lagrangian frame. 
\vskip 0.1cm

If the original process $\,x_t\,$ is stationary with $\,u^i(x)$, 
$\,D^{ij}(x,y)\,$ and the single-time PDF $\,\rho(x)\,$ time independent 
then the corresponding mean local velocity field $\,v(x)\,$ is also 
time-independent. \,The Lagrangian-frame process 
$\,\tilde x_t$, \,however, \,is non-stationary if $\,v\,$ does not
vanish, although its single-time PDF is equal to $\,\rho(\tilde x)\,$ and 
does not change in time. {\,\bf The stationary non-equilibrium dynamics 
becomes in the Lagrangian frame a non-stationary equilibrium one with
the same invariant probability density}. 
\vskip 0.1cm

In the case where the original process has time-independent instantaneous 
PDF's with vanishing probability current, the Eulerian and the Lagrangian 
frame processes coincide. However, for a non-equilibrium
Langevin dynamics
\qq
\dot{x}_t^i\ =\ -\beta\,d^{ij}_t(x_t)\,(\partial_jH_t)(x_t)\ +\ r^i_t(x_t)
\ +\ F^i_t(x)\ +\ \zeta^i_t(x)
\label{nonequileq}
\qqq
with a time-dependent Hamiltonian $\,H_t\,$ or/and an additional 
non-conservative force $\,F_t\,$ that generate non-trivial probability
current, the passage to the Lagrangian frame of mean local velocity
$\,v_t\,$ recasts the dynamics into the equilibrium form (\ref{tildequileq})
with a time-independent Hamiltonian and no non-conservative force.
The same is true for the process satisfying the equilibrium Langevin
equation (\ref{equileq}) but with non-Gibbsian instantaneous densities
(relaxing to equilibrium or not).
\vskip 0.2cm

\begin{figure}[t!]
\begin{center}
\vskip 0.2cm
\leavevmode
{%
\hspace*{0.4cm}
\begin{minipage}{0.4\textwidth}
        \includegraphics[width=5.5cm,height=7.2cm,angle=-90]{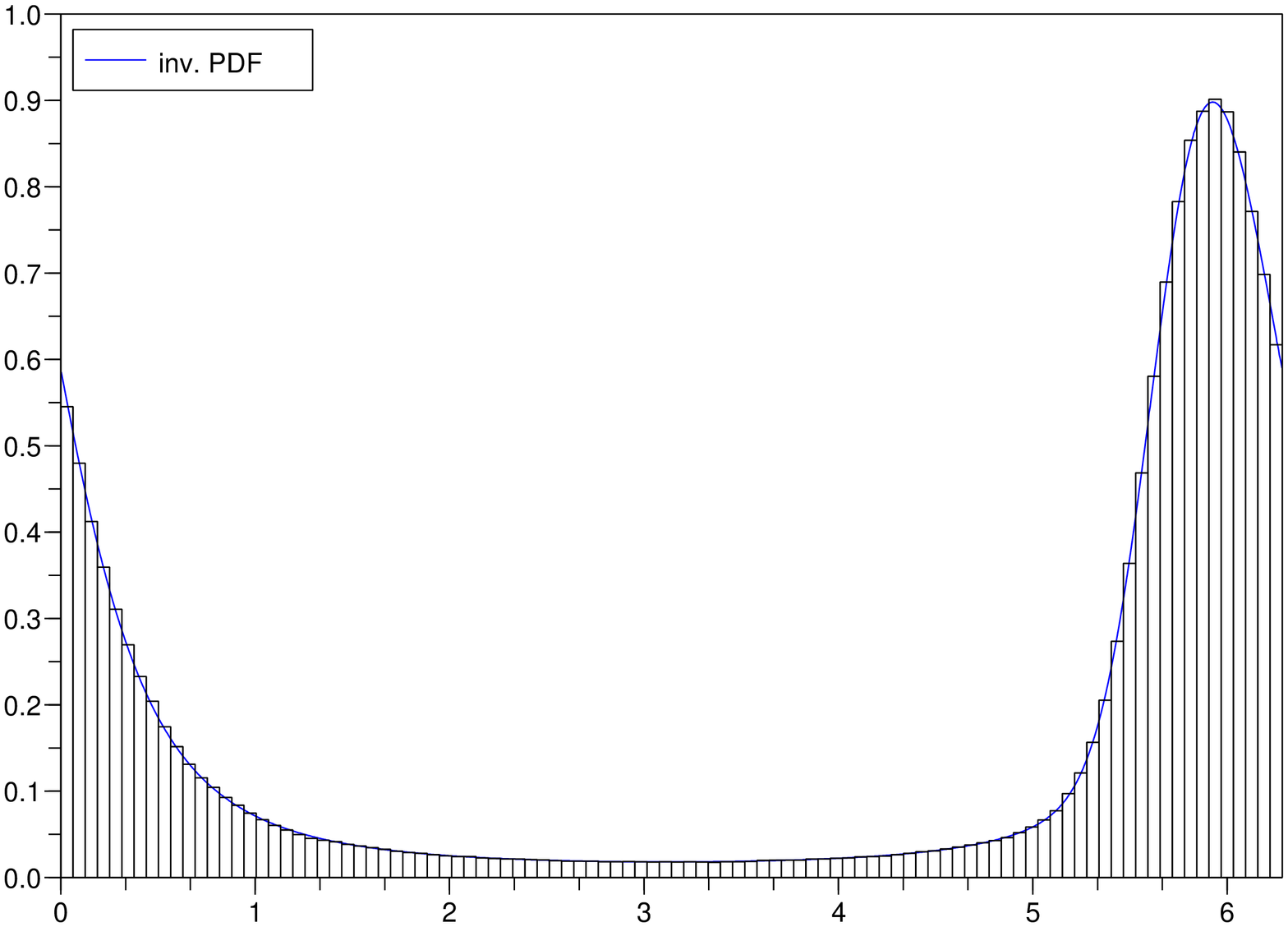}\\
        \vspace{-0.8cm} \strut
        \end{minipage}}
    \hspace{-4.8cm}
{%
      \begin{minipage}{0.2\textwidth}
        \includegraphics[width=2.1cm,height=3.2cm,angle=-90]{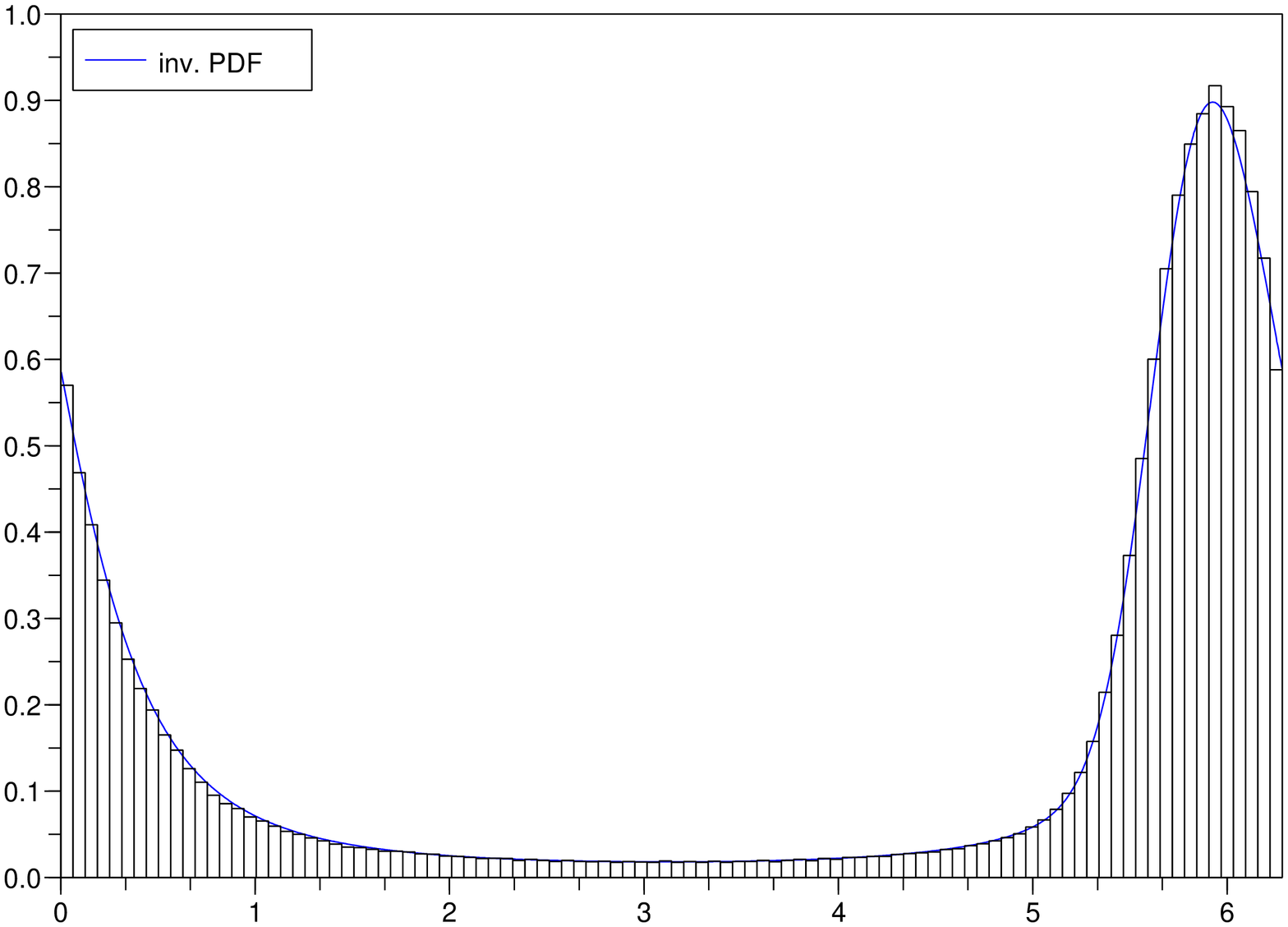}\\
      \vspace{1.1cm} \strut
        \end{minipage}}\hspace*{1.5cm}
{%
      \begin{minipage}{0.2\textwidth}
       \includegraphics[width=5.5cm,height=7.2cm,angle=-90]{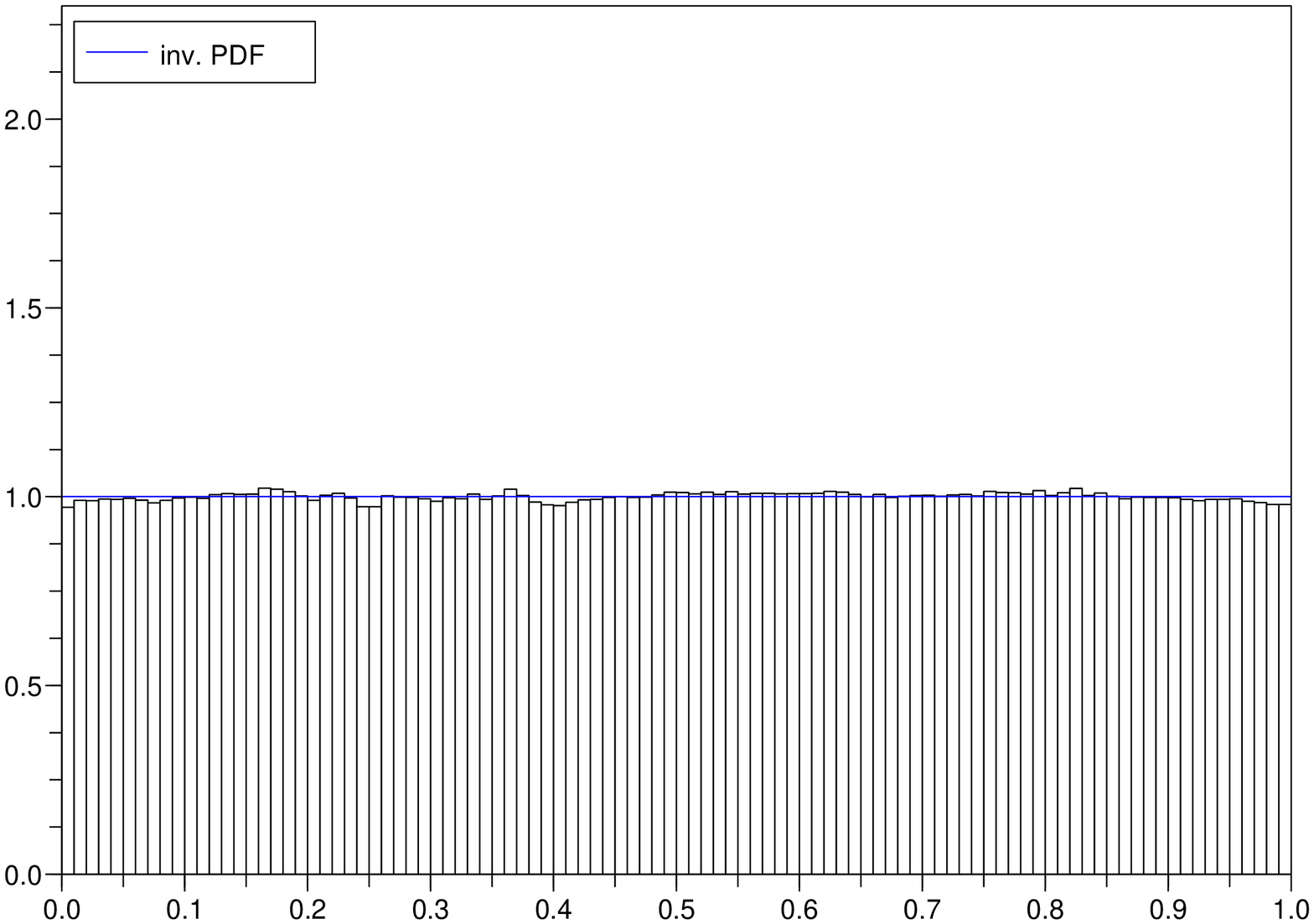}\\
      \vspace{-0.8cm} \strut
        \end{minipage}}
    \hspace{-1.4cm}
{%
      \begin{minipage}{0.2\textwidth}
        \includegraphics[width=2.1cm,height=3.2cm,angle=-90]{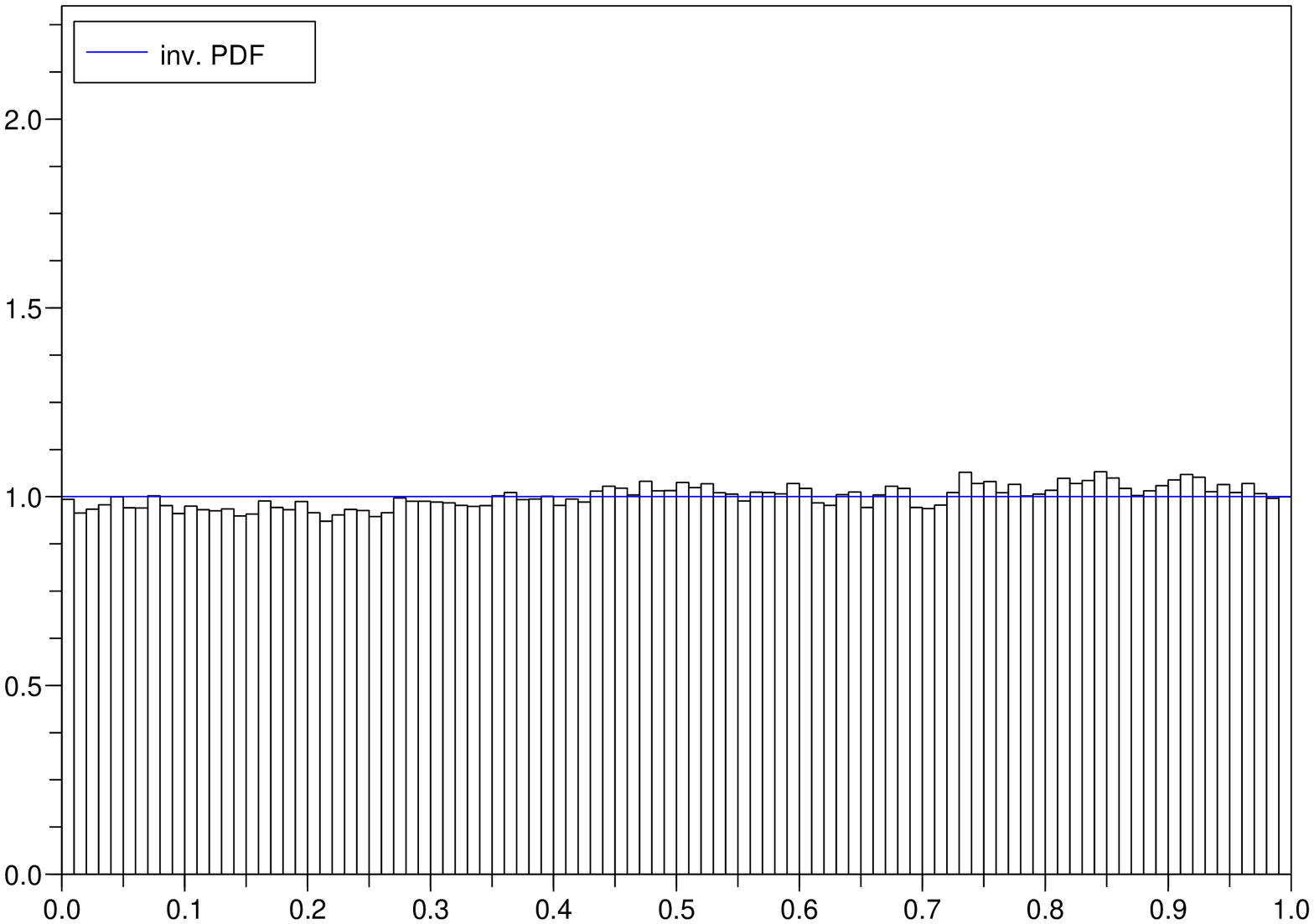}\\
      \vspace{1.1cm} \strut
        \end{minipage}}\hspace*{0.5cm}
\hspace*{2cm}
\end{center}
\hspace*{1.4cm}\parbox{13.5cm}{
\caption{\small{\ Left: \ \ \ theoretical invariant PDF $\,\rho(\theta)\,$ 
(blue solid line) 
\,compared to the histogram\break\hspace*{2.75cm}of 30000 time values 
on 1500 trajectories of the processes $\,\theta_t$. \ In the 
insert\break\hspace*{2.75cm}the same figure for $\,\tilde\theta_t\,$ 
undistinguishable with bare eye from the one for 
$\,x_t$\hfill\break\hspace*{1.6cm}\vspace{-0.3cm} 
\break\hspace*{1.6cm}Right: the same figures for the process $\,x_t\,$
obtained by the change of variables\break\hspace*{2.74cm}(\ref{chvar})}}}
\label{fig:fig0}
\end{figure}
\begin{figure}[t!]
\label{fig:fig2}
\begin{center}
\vskip -0.2cm
\leavevmode
{%
\hspace*{-1.8cm}
\begin{minipage}{0.4\textwidth}
        \includegraphics[width=5.5cm,height=7.2cm,angle=-90]{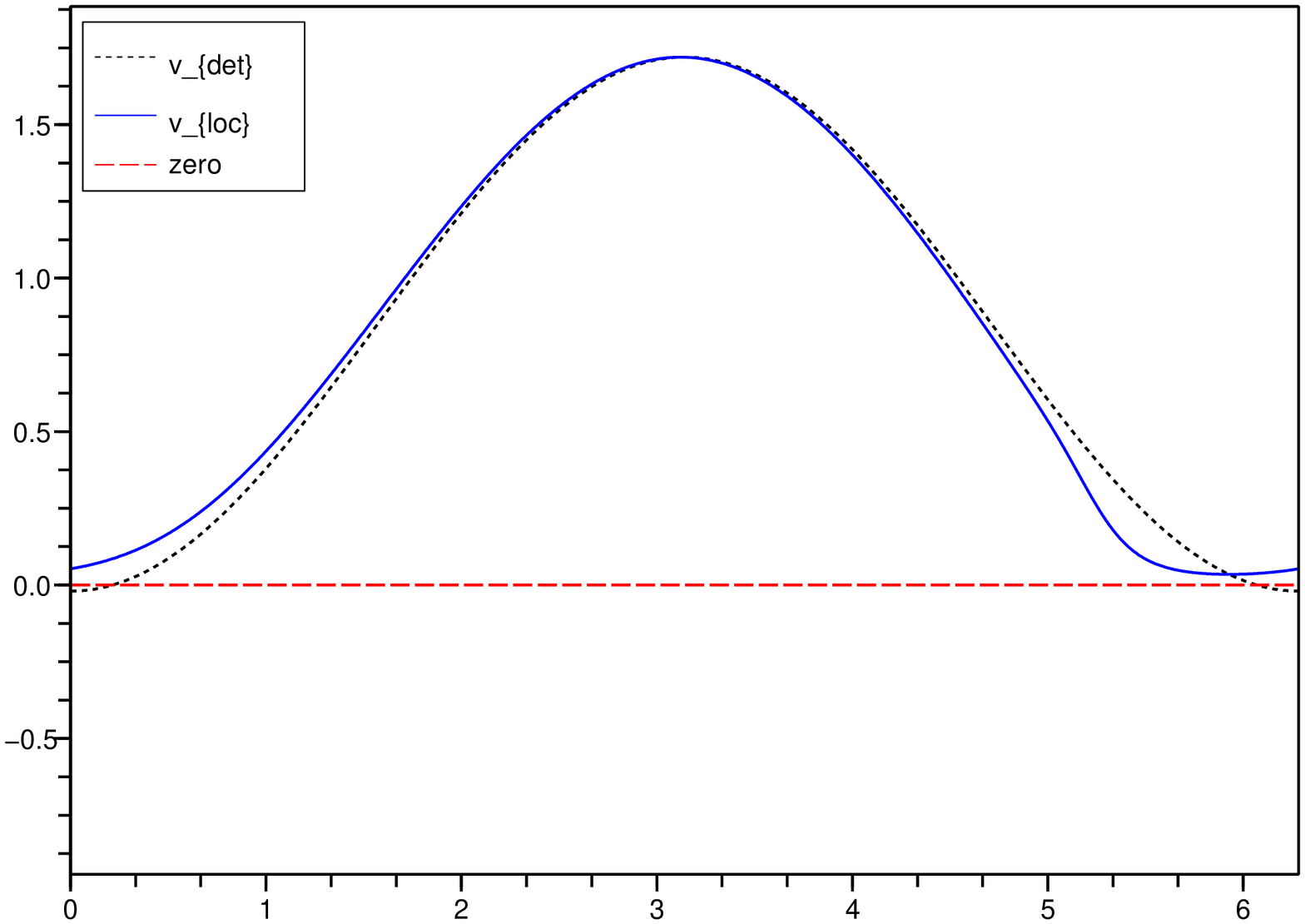}\\
        \vspace{-0.8cm} \strut
        \end{minipage}}
    \hspace{-5.85cm}
{%
      \begin{minipage}{0.2\textwidth}
        \includegraphics[width=1.3cm,height=2cm,angle=-90]{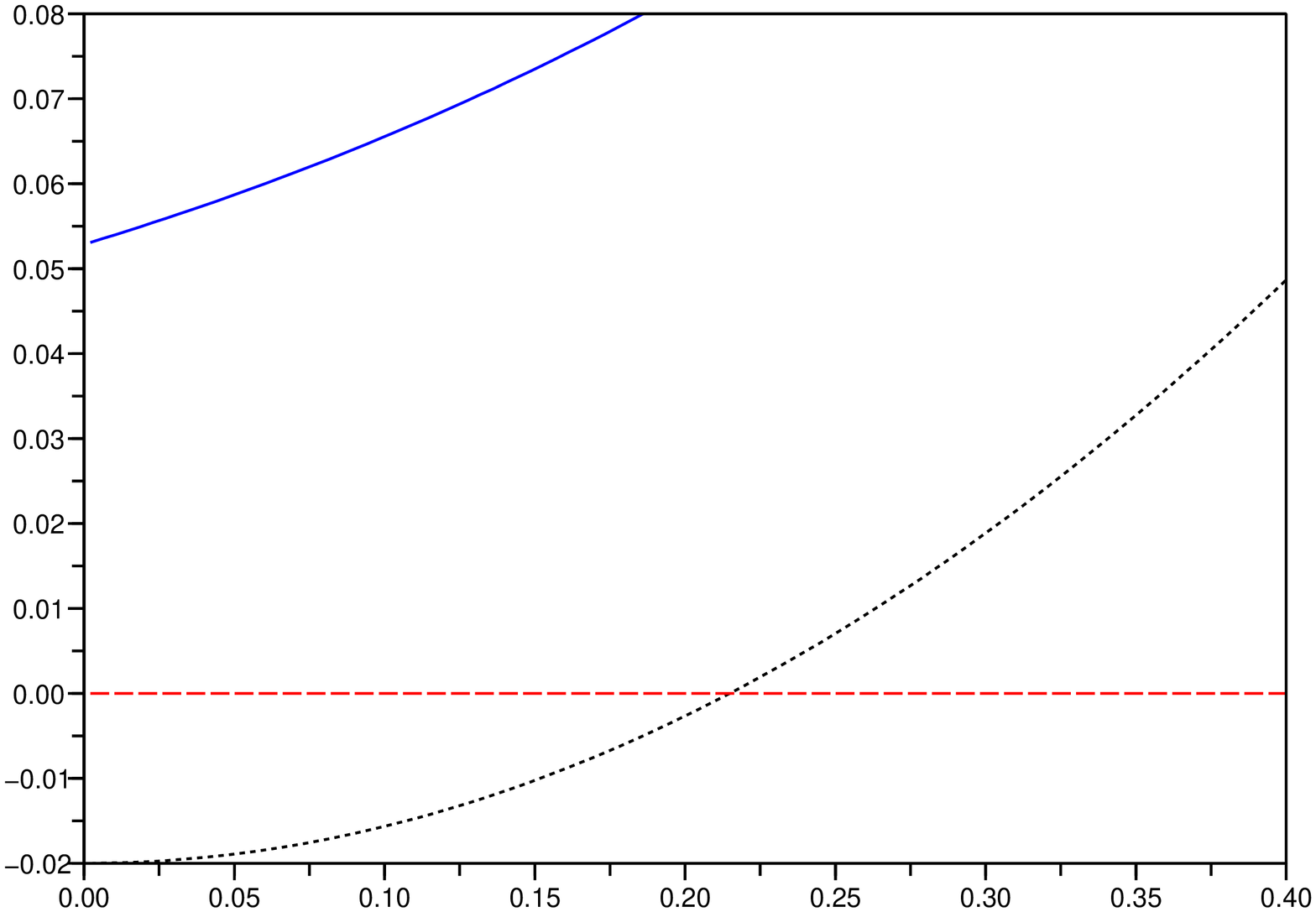}\\
      \vspace{-3.6cm} \strut
        \end{minipage}}\hspace*{-0.22cm}
{%
      \begin{minipage}{0.2\textwidth}
       \includegraphics[width=1.3cm,height=2cm,angle=-90]{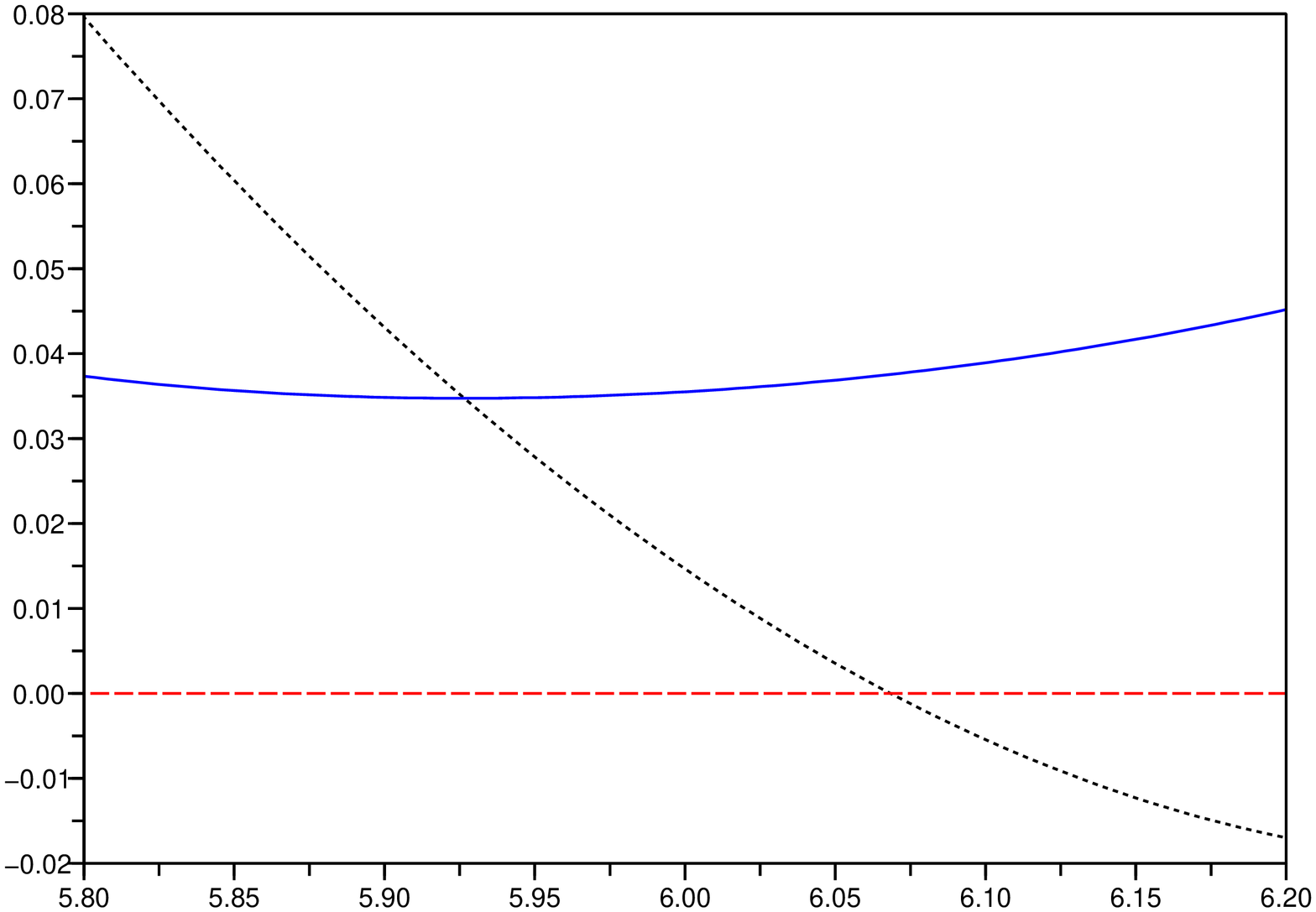}\\
      \vspace{-3.6cm} \strut
        \end{minipage}}
    \hspace{-0.9cm}
{%
      \begin{minipage}{0.2\textwidth}
        \includegraphics[width=5.5cm,height=7.2cm,angle=-90]{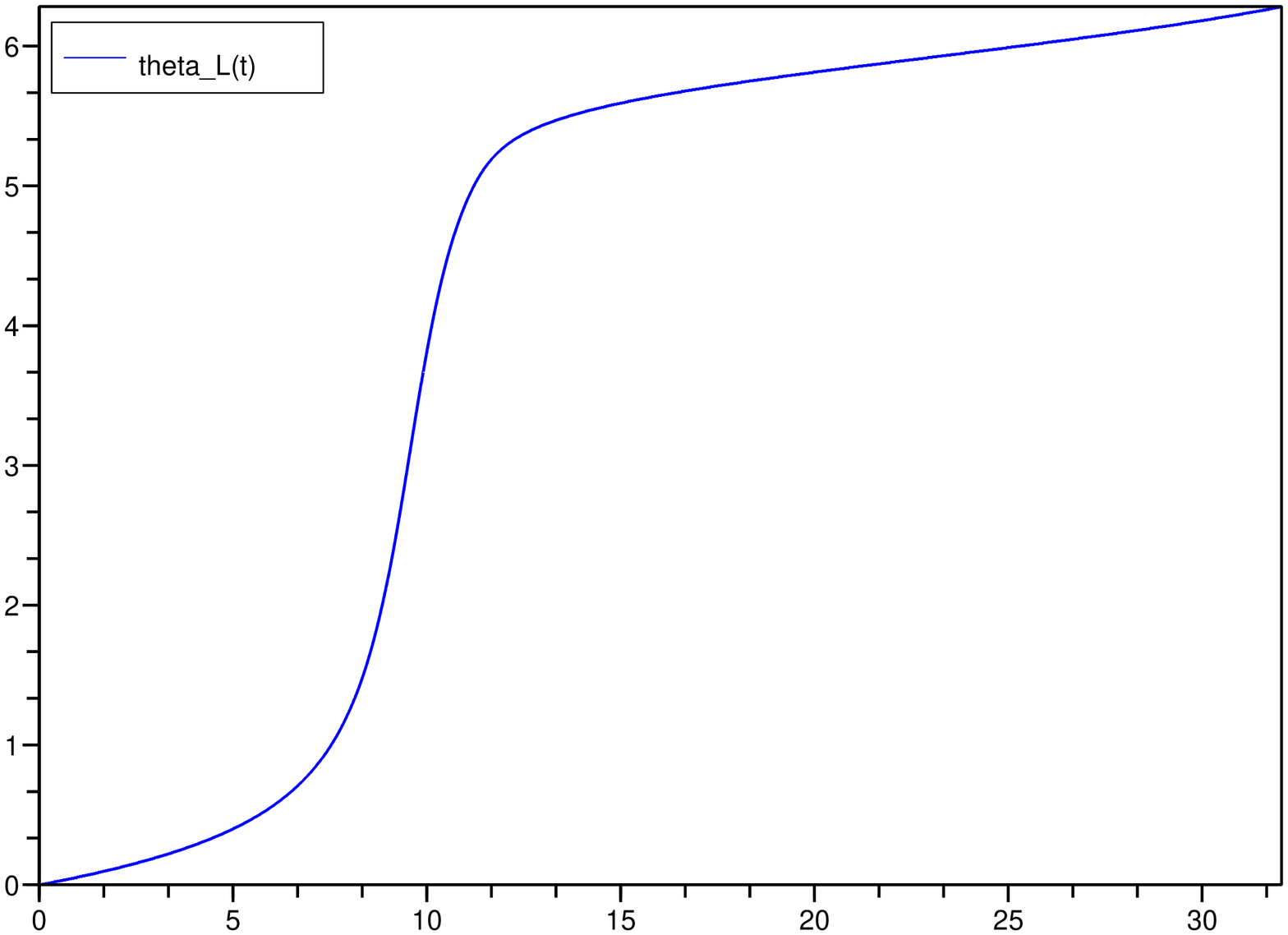}\\
      \vspace{-0.8cm} \strut
        \end{minipage}}\hspace*{0.5cm}
\hspace*{2cm}
\end{center}
\hspace*{1.4cm}\parbox{13.5cm}{
\caption{\small{Left: \,mean local velocity (blue solid line, 
everywhere positive) 
as compared\break\hspace*{2.84cm}to the deterministic velocity equal 
to the drift term in Eq.\,(\ref{1dSDE}) 
(black\break\hspace*{2.84cm}dotted line changing sign, with a repulsive and 
an attractive fixed points\break\hspace*{2.73cm} well visible in the 
blowups)\hfill\break\hspace*{1.7cm}\vspace{-0.3cm} 
\break\hspace*{1.7cm}Right: \ Lagrangian trajectory 
$\,\theta_L(t)\,$ of the mean local velocity with $\,\theta_L(0)=0$\hfill}}} 
\end{figure}
\vskip 0.5cm

\subsection{Examples}

\subsubsection{Colloidal particle on a circle}

The simplest example of a non-equilibrium Langevin dynamics
is provided by the overdamped motion of a particle
on a circle with its angular position satisfying the stochastic equation
\qq
\dot{\theta}_t\ =\ -(\partial_\theta H)(\theta_t)\,+\,F\,+\,\zeta_t
\label{1dSDE}
\qqq
\noindent with a periodic potential $\,H(\theta)=H(\theta+2\pi)$, \,a constant 
(non-conservative) force $\,F$, \,and a white noise $\,\zeta_t\,$ with
covariance $\ \langle\zeta_t\,\zeta_s\rangle=2D\delta(t-s)$.
\ Eq.\,(\ref{1dSDE}) has a stationary solution with
the invariant PDF $\,\rho\,$ given by the formula:
\qq
\rho(\theta)\ =\ Z^{-1}\ee^{-\frac{1}{D}(H(\theta)-F\theta)}\Big(
\int\limits_0^\theta\ee^{\,\frac{1}{D}(H(\vartheta)-F\vartheta)}d\vartheta\,
+\,\ee^{\,\frac{2\pi F}{D}}
\int\limits_x^{2\pi}\ee^{\,\frac{1}{D}(H(\vartheta)-F\vartheta)}d\vartheta\Big)\,,
\qqq 
where $\,Z\,$ is the normalization factor. \,The current corresponding
to this density is constant:
\qq
j\ =\ \big[-(\partial_\theta H)(\theta)+F-D\partial_\theta\big]\rho(\theta)\ 
=\ DZ^{-1}\big(\ee^{\,\frac{2\pi F}{D}}-1\big)\,.
\qqq
The one-dimensional dynamics becomes simpler in the variable 
\qq
x\ =\ \int\limits_0^\theta\rho(\vartheta)\,d\vartheta\,.
\label{chvar}
\qqq
taken modulo $\,1$. \,Note that $\,\frac{dx}{d\theta}
=\rho(\theta)=jv(\theta)^{-1}\,$ so that $\,j^{-1}x\,$ is the time
that the Lagrangian trajectory $\,\theta_L(t)\,$ of the mean local 
velocity starting at $\,\theta=0\,$ takes to get to $\,\theta$. 
In the variable $\,x$, \,the invariant
density $\,\rho(x)\equiv 1\,$ and Eq.\,(\ref{1dSDE}) takes
the form
\qq
\dot{x}_t\ =\ j\,+\,r(x_t)\,+\,\zeta_t(x_t)\,,
\qqq 
where $\ \zeta_t(x)=\rho(\theta)\zeta_t\ $ for $\ \theta=\theta_L(j^{-1}x)\ $
and
\qq
r(x)\ =\ D\rho(\theta)\,\partial_x\rho(\theta)\ =\ 
D(\partial_\theta\rho)(\theta)\ =\ 
\big[-(\partial_\theta H)(\theta)+F\big]\rho(\theta)\,-\,j\,.
\qqq
In the variable $\,x$, \,the mean local velocity $\,v(x)\equiv j$.
\,The corresponding Lagrangian-frame process
$\,\tilde x_t=x_t-j(t-t_0)\,$ and it satisfies the equilibrium-type
Langevin equation
\qq
\dot{\tilde x}_t\ =\ \tilde r_t(\tilde x_t)\,
+\,\tilde\zeta_t(\tilde x_t)
\qqq
with $\,\tilde r_t(\tilde x)=r(\tilde x+j(t-t_0))\,$ and 
$\,\tilde\zeta_t(\tilde x)=\zeta_t(\tilde x+j(t-t_0))\,$ and a constant
Hamiltonian.
\vskip 0.1cm

Fig.\,1 and Fig.\,2 represent the invariant density and the mean local 
velocity with its Lagrangian trajectory, both for the process 
$\,\theta_t\,$ satisfying Eq.\,(\ref{1dSDE}) with $\,H(\theta)
=0.87s^{-1}\times\sin(\theta)$, $\,F=0.85s^{-1}\,$ 
and $\,D=0.036s^{-1}$. \,Such process models the dynamics of a colloidal 
particle kept by an optical tweezer on a nearly circular orbit in the 
experiment described in \cite{GPCCG}.

\subsubsection{Linear stochastic equations}

A general class of explicitly soluble examples of non-equilibrium
dynamics, with multiple applications, is provided by stationary linear 
SDEs in $d$ dimensions of the form:
\qq
\dot{x}_t\ =\ Mx_t\,+\,\zeta_t\,,
\label{linear}
\qqq
where $\,M\,$ is a matrix whose eigenvalues have negative real part and
where
\qq
\big\langle\,\zeta_t^i\ \zeta_s^j\,\big\rangle\ =\ 2\,
D^{ij}\,\delta(t-s)\,,
\label{ncov}
\qqq
with a positive matrix $\,D=(D^{ij})$. \ Here, the invariant 
density has the Gaussian form \cite{CG1}
\qq
\rho(x)\ =\ Z^{-1}\ee^{-\,\beta\,H(x)}
\qqq
\vskip -0.4cm
\noindent with
\vskip -0.7cm
\qq
H(x)\ =\ \frac{_1}{^{2\beta}}\,x\cdot C^{-1}x\qquad{\rm for}\qquad
C\,=\,2\int\limits_0^\infty\ee^{tM}D\,\ee^{tM^T}\,dt\,.
\qqq
\vskip -0.3cm
\noindent The time-integral in the formula for the covariance $\,C\,$ 
converges due to the assumption on the eigenvalues of $\,M$. \,The mean 
local velocity corresponding to $\,\rho(x)\,$ is
\qq
v(x)\ =\ (M\,+\,DC^{-1})\,x
\label{locvlin}
\qqq
so that it depends linearly on $\,x$. \,The Lagrangian-frame process
\qq
\tilde x_t\ =\ \ee^{-(M+DC^{-1})(t-t_0)}x_t
\qqq
satisfies the time-dependent equilibrium-type linear
Langevin equation
\qq
\dot{\tilde x}_t\ =\ -\beta\,\tilde D_t\nabla H(\tilde x)\,+\,\tilde\zeta_t
\qqq
where the white noise $\ \tilde\zeta_t\,=\,\ee^{-(M+DC^{-1})(t-t_0)}\zeta_t\ $ 
has the covariance
\qq
\big\langle\,\tilde\zeta_t^i\,\tilde\zeta_s^j\,\rangle\ =\ 2\,\delta(t-s)\,
\tilde D^{ij}_t\qquad{\rm with}\qquad\tilde D_t\ =\ \ee^{-(M+DC^{-1})(t-t_0)}
\,D\hspace{0.075cm}\ee^{-(M+DC^{-1})^T(t-t_0)}\,.
\qqq

\subsubsection{Sheared suspensions}
\label{sec:shear}

Stochastic equations of the type (\ref{SDE}) may be used to
model the dynamics of suspensions of colloidal particles \cite{Hunter} 
or of a polymer, undergoing an overdamped motion driven by conservative 
forces and opposed by friction, see \cite{SpS3} for a recent
discussion. An example is provided by the set of equations for 
the three-dimensional positions $\,\bm r_i\,$ of $\,N\,$ particles:
\qq
\gamma\,\dot{r}^a_i\,=\,-\,\partial_{r^a_i} H(\underline{\bm r})\,
-\,\gamma\,u^a_t(\bm r_i)\,+\,\zeta^a_{i,t}\,,
\label{susp}
\qqq
where $\,\gamma\,$ is the friction coefficient,
$\ \underline{\bm r}=(\bm r_i)_{i=1}^N$, 
$\ H(\underline{\bm r})\,$ is the potential
energy and $\,\bm u(t,\bm r)\,$ is the velocity field of the solvent.
$\,\xi^a_{i,t}\,$ are the components of the white noise with the covariance
\qq
\big\langle\,\zeta^a_{i,t}\,\zeta^b_{j,s}\,\big\rangle\ =\ 2\,\gamma\,\beta^{-1}
\delta^{ab}\,\delta_{ij}\,\delta(t-s)\,.
\qqq  
For a diluted colloidal suspension, assuming only 2-body isotropic interactions,
one may take
\qq
H(\underline{\bm r})\ =\ \sum\limits_{i<j}U(r_{ij})\,+\,
\sum\limits_iU_0({\bm r}_i)
\qqq
for $\,r_{ij}\equiv|{\bm r}_i-{\bm r}_j|\,$
and for the polymer modeled as a chain of beads with nearest neighbor
interaction (Rouse model \cite{Rouse}),
\qq
H(\underline{\bm r})\ =\sum\limits_{i<N}
U(r_{i\,(i+1)})\,+\,\sum\limits_iU_0({\bm r}_i)\,.
\qqq
If the solvent is at rest, and the external potential $\,U_0\,$ is confining
then the detailed balance holds for the normalized Gibbs density 
$\,\rho_0(\underline{\bm r})=Z^{-1}\ee^{-\beta H(\underline{\bm r})}\,$ which
is left invariant under evolution. \,If, however, the 
solvent undergoes a shear flow with $\,{\bm u}_t(\bm r)=
f({\bm r}\cdot{\bm e}_1)\,{\bm e}_2$,
\,where $\,{\bm e}_i\,$ are the vectors of the canonical basis of $\,\NR^3$, 
\,or a vortical motion with $\,\bm u_t(\bm r)
=g(|{\bm r}\wedge{\bm e}_3|)\,{\bm r}\wedge{\bm e}_3$, \,then the detailed 
balance is broken and the mean local velocity becomes equal in the stationary
state to
\qq
v^a_i(\underline{\bm r})\ =\ -\,\gamma^{-1}\partial_{r^a_i}H(\underline{\bm r})
\,-\,u^a_t(\bm r_i)\,-\,(\gamma\beta)^{-1}
\,\partial_{r^a_i}\ln\rho(\underline{\bm r})\,.
\qqq
where $\,\rho(\underline{\bm r})\,$ 
is the non-Gibbsian invariant density. 
\vskip 0.1cm

In general, the form of $\,\rho(\underline{\bm r})\,$ 
is difficult to access in realistic situations. One of exceptions is the 
idealized case leading to the linear stochastic equations describing the 
simplest realization of the Rouse model of a polymer suspension with 
$\,U(r)=\frac{1}{2}\kappa r^2$ and $\,U_0(\bm r)=\frac{1}{2}k r^2\,$ and with 
linear velocity field $\,\bm u_t(\bm r)$. \,For the vortical velocity
$\,\bm u_t(\bm r)=\omega\,{\bm r}\wedge{\bm e_3}$, \,the stochastic 
equation (\ref{susp}) takes the form (\ref{linear}) with  
\qq
M^{ab}_{ij}\,=\,-\gamma^{-1}\delta^{ab}(-\kappa\Delta_{ij}+k
\delta_{ij})\,+\,\omega\,\epsilon_{ab3}\,\delta_{ij}\,,
\qqq  
where $\,\Delta_{ij}=\hspace{-0.2cm}\sum\limits_{|i'-i|=1}
\hspace{-0.2cm}(\delta_{i'j}-\delta_{ij})$, 
\,\,and with the matrix
\vskip -0.3cm 
\qq
D^{ab}_{ij}\,=\,(\gamma\beta)^{-1}\delta^{ab}\,\delta_{ij}\,.
\qqq
in the noise covariance (\ref{ncov}).
\,In spite of the vortical motion of the solvent, the Gibbs density
$\,\rho_0(\underline{\bm r})\,$ independent of the vorticity $\,\omega\,$ 
remains invariant for the symmetry reasons. \,Nevertheless, for 
$\,\omega\not=0$, \,the detailed balance is broken
and the mean local velocity is given by the solvent velocity  
\qq
v^a_i(\underline{\bm r})\ =\ \omega\,{\bm r}_i\wedge{\bm e_3}\,.
\qqq
The Lagrangian frame just rigidly rotates around the third axis with the 
angular velocity $\,\bm\Omega=\omega\,{\bm e}_3\,$ and the Lagrangian-frame
process $\,\tilde{\bm r}_{i,t}\,$ satisfies the stochastic equation
(\ref{susp}) with $\,{\bm u}_t\,$ set to zero.
\vskip 0.1cm

Keeping the same harmonic potentials but replacing the vortical solvent motion
by the shear flow with $\,u_t(\bm r)\,=\,s\,({\bm r}\cdot{\bm e}_1)\,
{\bm e}_2\,$ with a constant shear rate $\,s$, \,we obtain the linear 
stochastic equation (\ref{linear}) with
\qq
M^{ab}_{ij}\ =\ -\gamma^{-1}\delta^{ab}(-\kappa\Delta_{ij}+k
\delta_{ij})\,+\,s\,\delta^{a2}\delta^{b1}\,\delta_{ij}
\qqq  
and the noise covariance as before. 
\,The $\,N\times N\,$ matrix $\,-\Delta=(-\Delta_{ij})\,$ has 
the eigenvalues $\,\omega_k=2\big[1-\cos\big(\frac{\pi k}{N}\big)\big]\,$ 
corresponding to the normalized eigenvectors
\qq
(\varphi^\ell_j)\,=\,\big((\frac{_2}{^N})^{1/2}
\cos\big(\frac{_{\pi\ell(j-\frac{_1}{^2})}}{^N}\big)
\big)\qquad{\rm for}\qquad\ell=0,1,\dots,N-1\,.
\qqq
The passage to the Fourier modes $\,\hat{\bm r}_\ell\equiv{\bm r}_j
\varphi_j^\ell\,$ (sum over $j$) diagonalizes matrix 
$\,M\,$ into $\,3\times3\,$ blocs
with the entries $\,M^{ab}_\ell=\delta^{ab}\mu_\ell+s\,
\delta^{a2}\delta^{b1}\,$ for $\,\mu_\ell\equiv\kappa\omega_\ell+k$.
\,The invariant density $\,\rho(\underline{\bm r})\,$ is Gaussian. Its
covariance depends quadratically on the shearing rate $\,s\,$ 
and is composed of the $\,3\times3\,$ blocs
\qq
&&C^{ab}_\ell\ =\ \frac{_1}{^{\beta\mu_\ell}}\big[\delta^{ab}\,+\,
\sigma_\ell(\delta^{a1}\delta^{2b}+\delta^{a2}\delta^{1b})
\,+\,2\sigma_\ell^2\delta^{a2}\delta^{2b}\big],
\qqq
for $\,\sigma_\ell\equiv\frac{{\gamma s}}{{2\mu_\ell}}=\frac{\gamma s}
{2(\kappa\omega_\ell+k)}$, \,with the 
blocs of the inverse covariance:
\qq
(C^{-1})^{ab}_\ell\ =\ 
\beta\mu_\ell\Big[\delta^{ab}\,+\,
\frac{_{\sigma_\ell^2}}{^{1+\sigma_\ell^2}}\,
(\delta^{a1}\delta^{1b}-\delta^{a2}\delta^{2b})\,-\,
\frac{_{\sigma_\ell}}{^{1+\sigma_\ell^2}}\,
(\delta^{a1}\delta^{2b}+\delta^{a2}\delta^{1b})\Big].
\qqq
The mean local velocity has the Fourier components
\qq
\hat v(\underline{\bm r})^a_\ell\ =\ (M+DC^{-1})^{ab}_\ell\,{\hat r}^b_\ell\,,
\qqq
see Eq.\,(\ref{locvlin}), with
\qq
(M+DC^{-1})^{ab}_\ell\ =\ \frac{_{\mu_\ell}}{^{\gamma}}
\Big[\frac{_{\sigma_\ell^2}}{^{1+\sigma_\ell^2}}\,
(\delta^{a1}\delta^{1b}-\delta^{a2}\delta^{2b})\,-\,
\frac{_{\sigma_\ell}}{^{1+\sigma_\ell^2}}\,
(\delta^{a1}\delta^{2b}+\delta^{a2}\delta^{1b})\,
+\,2\sigma_\ell\delta^{a2}\delta^{1b}\Big]
\qqq
and is incompressible. Its Lagrangian flow is linear.
It factorizes for different Fourier modes and takes place
along ellipses in the planes orthogonal to $\,{\bm e}_3$:
\qq
\hat\phi_t(\underline{\bm r})^a_\ell
&=&\delta^{a3}\hat r^3_\ell\,+\,
(\delta^{a1}\hat r^1_\ell+\delta^{a2}\hat r^2_\ell)\,
\cos\Big(\frac{_{s(t-t_0)}}{^{2\sqrt{1+\sigma_\ell^2}}}\Big)
\,+\,\Big[\,(\delta^{a1}\hat r^1_\ell-\delta^{a2}\hat r^2_\ell)\,
\frac{_{\sigma_\ell}}{^{\sqrt{1+\sigma_\ell^2}}}\cr
&&-\,(\delta^{a1}\hat r^2_\ell+\delta^{a2}\hat r^1_\ell)\,
\frac{_{1}}{^{\sqrt{1+\sigma_\ell^2}}}\,+\,2\delta^{a2}\hat r^1_\ell
\,\sqrt{1+\sigma_\ell^2}\,\Big]\,
\sin\Big(\frac{_{s(t-t_0)}}{^{2\sqrt{1+\sigma_\ell}}}\Big),
\qqq
see Fig.\,3. \,The ellipses are more and more elongated in the direction
of the flow with increasing shearing rate $\,s\,$ and decreasing 
Fourier mode $\,\ell$.  
\begin{figure}[t!]
\begin{center}
\vskip 0.2cm
\leavevmode
        \includegraphics[width=6.1cm,height=8.2cm,angle=-90]{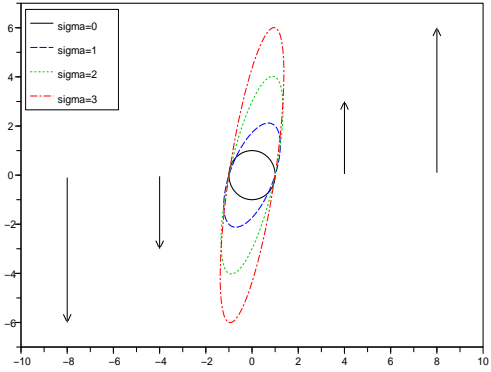}
\end{center}
\hspace*{1.4cm}\parbox{13.5cm}{
\caption{\small{\ \,Ellipses followed in the $xy$-plane under the Lagrangian 
flow of mean local velocity\break\hspace*{1.72cm}by the Fourier modes 
$\,\hat{\bm r}_\ell\,$ starting at $\,(1,0)$ for different values of 
the parameter\break\hspace*{1.72cm}$\sigma_\ell=\frac{\gamma s}{4\kappa
\big(1-cos(\frac{\pi\ell}{N})\big)+2k}$}}}
\label{fig:ellipse}
\end{figure}
The formula for the time-dependent covariance $\,\tilde D_t\,$ of the
noise in the Lagrangian-frame process is given in Appendix C.
\vskip 0.1cm

\subsection{Back to Eulerian frame}

When passing to the Lagrangian frame, a part of the information
about the system contained in the probability current or mean local 
velocity is lost. If we want to reconstruct the original 
Eulerian process $\,x_t$, \,we have to supply the forgotten information.
A convenient way to do that is to provide the local velocity
transformed to the Lagrangian frame:
\qq
\tilde v^i_t(\tilde x)\ =\ (\partial_k\Phi^{-1}_t)^i(x)\,v^k_t(x)\ =\ 
-(\partial_t\Phi^{-1}_t)^i(x)
\label{tocom1}
\qqq
for $\,x=\Phi_t(\tilde x)$. \,Given the vector field $\,\tilde v_t(\tilde x)$,
\,consider the flow of transformations $\ x\mapsto\tilde\Phi_t(x)\ $ 
such that
\qq
\partial_t\tilde\Phi_t(x)\ =\ -\tilde v_t
\big(\tilde\Phi_t(x)\big)\,,\qquad\tilde\Phi_{t_0}(x)\ =\ x\,.
\label{tocom2}
\qqq
The comparison of Eqs. (\ref{tocom1}) and (\ref{tocom2}) shows that
$\,\tilde\Phi_t(x)=\Phi_t^{-1}(x)$. \ This permits to reconstruct the original
process as
\qq
x_t\ =\ \tilde\Phi^{-1}_t(\tilde x_t)
\qqq
and the original mean local velocity as
\qq
v^i_t(x)\ =\ (\partial_j\tilde\Phi^{-1}_t)^i(\tilde x)\,\tilde v^j_t
(\tilde x)\ =\ -(\partial_t\tilde\Phi^{-1}_t)^i(\tilde x)
\qqq
for $\,\tilde x=\tilde\Phi_t(x)$. \ In the special case when 
the mean local velocity is time-independent
(for example when $\,x_t\,$ is a stationary process),
\qq
\tilde v^i_t(\tilde x)\ =\ -(\partial_t\Phi^{-1}_t)^i(x)\ =\ 
-(\partial_t\Phi_{-t})^i(x)\ =\ v^i\big(\Phi_{-t})^i(x)\big)\ =\ v^i(\tilde x)
\qqq
so that the velocity field $\,\tilde v_t\,$ coincides with the
mean local velocity of the Eulerian frame and is time-independent.  
\vskip 0.1cm

As we see, the knowledge of the non-equilibrium diffusion $\,x_t\,$
is equivalent to the knowledge of the equilbrium diffusion $\,\tilde x_t\,$
and of the (deterministic) velocity field $\,\tilde v_t$.

\nsection{Diffusion Processes with Hamiltonian forces}
\label{sec:Langev}

\subsection{Modified probability current and mean local velocity}
\label{sec:modcurr}

In many applications, one deals with non-equilibrium diffusions
in the presence of Hamiltonian forces. It is then useful
to single out their contribution and to replace the SDE (\ref{SDE}) by
\qq
\dot{x}_t^i&=&\,u^i_t(x)\,+\,\Pi^{ij}_t(x_t)\,(\partial_jH)(x_t)\,
-\,\beta^{-1}\,(\partial_j\Pi^{ij}_t)(x_t)\,+\,\zeta^i_t(x_t)
\label{nonequileq11}
\qqq
where the term $\,\Pi^{ij}_t\,\partial_jH_t\,$ with 
$\,\Pi_t^{ij}=-\Pi_t^{ji}\,$ stands for the Hamiltonian force. Geometrically,
the antisymmetic tensor field $\,\Pi_t^{ij}\,$ represents a (possibly
time dependent) Poisson structure but we shall not need its property that 
assures the Jacobi identity of the Poisson bracket. The subtraction of 
$\,\beta^{-1}\partial_j\Pi^{ij}_t\,$ on the right hand side of 
Eq.\,(\ref{nonequileq11}) assures that the terms involving $\,\Pi_t\,$ 
transform as a vector field under a change of coordinates if the Gibbs 
factor $\,\ee^{-\beta H}\,$ transforms as a density. \,An example of dynamics 
(\ref{nonequileq11}) is provided by the Langevin equation
\qq
\dot{x}_t^i&=&-\,\beta\,d^{ij}_t(x_t)\,(\partial_jH_t)(x_t)\,+\,r^i_t(x_t)\,
+\,F^i_t(x_t)\cr\cr
&&+\,\Pi^{ij}_t(x_t)\,(\partial_jH)(x_t)\,
-\,\beta^{-1}\,(\partial_j\Pi^{ij}_t)(x_t)\,+\,\zeta^i_t(x_t)\,,
\label{nonequileq1}
\qqq
compare to Eq.\,(\ref{nonequileq}). \,In the presence of Hamiltonian forces, 
it is convenient to redefine the probability current as
\qq
j^i_t\ =\ \big[\hat u^i_t\,+\,\Pi^{ij}_t(\partial_jH_t)\,-
\,b^{ij}_t\,\partial_j\big]\hspace{0.03cm}\rho_t\,,
\label{curr1}
\qqq
where $\,b^{ij}_t=d^{ij}_t-\beta^{-1}\Pi^{ij}_t$. \,The new expression
for the current $\,j_t\,$ differs from the one prescribed by Eq.\,(\ref{curr})
by the addition of the term $\ \beta^{-1}\partial_j\big(\Pi^{ij}_t\rho_t)$.
\,The continuity equation (\ref{conteq}) 
still holds  since the added term is divergence-less so that
the flux of $\,j_t\,$ through the boundary of any region $\,\CV\,$
still gives the rate of change of the probability that $\,x_t\,$
belongs to $\,\CV$. \,For the case of the Langevin equation 
(\ref{nonequileq1}), the new expression for the current reduces to
\qq
j^i_t\ =\ \big[-\beta\,b^{ij}_t\,\partial_jH_t\,+\,F^i_t\,
-\,b^{ij}_t\,\partial_j\big]\hspace{0.03cm}\rho_t\,.
\label{curr2}
\qqq
In particular, if $\,H_t\equiv H\,$ is time-independent and the 
additional force $\,F_t\equiv 0\,$ then the modified probability 
current (\ref{curr2}) associated to the Gibbs density 
$\,\rho(x)=Z^{-1}\ee^{-\beta\,H(x)}\,$ vanishes and $\,\rho\,$ is 
preserved by the evolution. It is then natural to extend the notion 
of equilibrium dynamics to such a case.
\vskip 0.1cm

As before, we may introduce the velocity field by the relation
\qq
v^i_t\ =\ \rho_t^{-1}j^i_t\ =\ \hat u^i_t\,+\,\Pi^{ij}_t(\partial_jH_t)\,
-\,b^{ij}_t\,\partial_j\ln{\rho_t}\,.
\label{locvels}
\qqq
Since now
\qq
v^i_t(x)\ =\ \frac{\big\langle\,\dot{x}_t\,\,
\delta(x-x_t)\,\big\rangle}{\big\langle\,\delta(x-x_t)\,\big\rangle}
\ +\ \beta^{-1}\rho_t(x)^{-1}\partial_j\big(\Pi^{ij}_t\rho_t\big)(x)\,,
\qqq
we shall call $\,v_t(x)\,$ the subtracted mean local velocity.
The continuity equation (\ref{conteq}) still takes the form of
the advection equation (\ref{adveq}).
\vskip 0.1cm

If we realize the passage to the Lagrangian frame of the velocity
$\,v_t\,$ of Eq.\,(\ref{locvels}) as described in Sect.\,\ref{sec:Lagrfr},
using  the flow of $\,v_t\,$ that we shall still denote by 
$\,\Phi_t\,$ and introducing the Lagrangian-frame process 
$\,\tilde x_t=\Phi_t^{-1}(x_t)$, 
then the considerations of Sect.\,\ref{sec:instdens}
go unchanged because they only use the advection equation (\ref{conteq}),
not the explicit form of $\,v_t(x)$. \,As before, we infer that 
the instantaneous PDF of the process $\,\tilde x_t\,$ is frozen to 
the time $\,t_0\,$ value $\,\rho_{t_0}\,$ of the PDF of the Eulerian process
$\,x_t$.  
\vskip 0.1cm 

On the other hand, in the derivation of the SDE for the Lagrangian-frame 
process in Sect.\,\ref{sec:Lfstocheq}, the explicit form of $\,v_t(x)\,$ 
was used in Eq.\,(\ref{explfor}). \,As a consequence, the SDE for 
$\,\dot{\tilde x}^i_t\,$ will pick now the additional term 
\qq
&&\hspace{-1cm}-(\partial_k\Phi^{-1}_t)^i(x_t)
\,\beta^{-1}\rho_t(x)^{-1}\partial_l
\big(\Pi^{kl}_t\,\rho_t\big)(x_t)\cr\cr 
&&\hspace{-1cm}=\ -\beta^{-1}(\partial_k\Phi^{-1}_t)^i(x_t)
\,\big[\Pi^{kl}_t(x_t)\,
(\partial_l\ln{\rho_t})(x_t)\,+\,(\partial_l\Pi^{kl}_t)(x_t)\big]\cr\cr
&&\hspace{-1cm}=\ -\beta^{-1}(\partial_k\Phi^{-1}_t)^i(x_t)
\,\big[\Pi^{kl}_t(x_t)\,(\partial_l\Phi^{-1}_t)^j(x_t)\,
(\partial_j\rho_{t_0})(\tilde x_t)\cr\cr
&&\hspace{2.3cm}+\,\Pi^{kl}_t(x_t)\,(\partial_j\Phi_t)^h(\tilde x_t)
\,(\partial_l\partial_h\Phi_t^{-1})^j(x_t)\,
+\,(\partial_l\Pi^{kl}_t)(x_t)\big]\,,
\label{corrr}
\qqq
where the second equality follows from Eq.\,(\ref{za:ln}) and the
identity $\,\tilde\rho_t=\rho_{t_0}$.
\ Introducing the Lagrangian-frame antisymmetric tensor field
\qq
\tilde\Pi^{ij}_t(\tilde x)\ =\ (\partial_k\Phi^{-1}_t)^i(x)\,\Pi^{kl}_t(x)\,
(\partial_l\Phi^{-1}_t)^j(x)
\qqq
where $\,x=\Phi_t(\tilde x)\,$ and observing that
\qq
(\partial_j\tilde\Pi^{ij}_t)(\tilde x)&=&\big[(\partial_h\partial_k
\Phi^{-1}_t)^i(x)\,\Pi^{kl}_t(x)\,(\partial_l\Phi^{-1}_t)^j(x)\,+\,
(\partial_k\Phi^{-1}_t)^i(x)\,(\partial_h\Pi^{kl}_t)(x)\,
(\partial_l\Phi^{-1}_t)^j(x)\cr\cr
&&\hspace{4cm}+\,
(\partial_k\Phi^{-1}_t)^i(x)\,\Pi^{kl}_t(x)\,
(\partial_h\partial_l\Phi^{-1}_t)^j(x)\big]\,(\partial_j\Phi_t)^h(\tilde x)
\cr\cr
&=&(\partial_k\Phi^{-1}_t)^i(x)\,(\partial_l\Pi^{kl}_t)(x)\,+\,
(\partial_k\Phi^{-1}_t)^i(x)\,\Pi^{kl}_t(x)\,
(\partial_h\partial_l\Phi^{-1}_t)^j(x)\,(\partial_j\Phi_t)^h(\tilde x)\,,
\qqq
we may rewrite the additional term (\ref{corrr}) as
\qq
-\beta^{-1}\tilde\Pi^{ij}_t(\tilde x_t)\,(\partial_j\rho_{t_0})(\tilde x_t)
\,-\,\beta^{-1}(\partial_j\tilde\Pi^{ij}_t)(\tilde x_t)
\ =\ \tilde\Pi^{ij}_t(\tilde x_t)\,\partial_j\tilde H(\tilde x_t)
\,-\,\beta^{-1}(\partial_j\tilde\Pi^{ij}_t)(\tilde x_t)\,.
\qqq
Altogether, the Lagrangian-frame process $\,\tilde x_t\,$ satisfies now the
equilibrium-type time-dependent SDE with a Hamiltonian force:
\qq
\dot{\tilde x}^i_t\ =\ -\beta\,\tilde d^{ij}_t(\tilde x_t)\,(\partial_j\tilde H)
(\tilde x_t)\,dt\,+\,
\tilde\Pi^{ij}_t(\tilde x_t)\,(\partial_j\tilde H)(\tilde x_t)
\,-\,\beta^{-1}(\partial_j\tilde\Pi^{ij}_t)(\tilde x_t)
\,+\,\tilde r^i_t(\tilde x_t)\,+\,\tilde\zeta^i_t(\tilde x_t)\,,
\label{equileq1}
\qqq 
Clearly, the modified probability current associated with the
conserved density $\,\rho_{t_0}=\tilde Z^{-1}\ee^{-\beta\,\tilde H}\,$ 
vanishes for the Lagrangian-frame process.

\subsection{Example of Langevin-Kramers dynamics}

The particular case of Langevin dynamics with Hamiltonian forces 
is provided by the $2^{\rm nd}$ order Langevin-Kramers SDE 
\qq
m_{ij}\,\ddot{q}^j_t\ =\ -\,\gamma_{ij}\,\dot{q}_t^j\,-\,\partial_iV_t(q_t)\,
+\,f_i(q_t)\,+\,\xi_{t,i}
\label{2nde}
\qqq
with the positive mass $\,m=(m_{ij})$ and friction $\,\gamma=(\gamma_{ij})\,$
matrices that, for simplicity, we assume independent of $\,t\,$ and 
$\,q$, \,with a potential $\,V_t(q)$ and a non-conservative force 
$\,f_t(q)$, \,and with a white noise $\,\xi_{t}\,$ with the covariance 
\qq
\langle\,\xi_{t,i}\,\xi_{s,j}\,\rangle\ 
=\ 2\,\beta^{-1}\sigma_{ij}\,\delta(t-s)\,.
\qqq
We keep the matrix $\,\sigma\,$ different from $\,\gamma\,$ to allow
noises modeling environments with variable temperature that violate
the Einstein relation $\,\sigma=\gamma$.
\,The $2^{\rm nd}$ order equation (\ref{2nde}) may be rewritten 
as the $1^{\rm st}$ order SDE (\ref{nonequileq1}) in the phase space 
of points $\,x=(q,p)\,$ if we set
\qq
&d\,=\,\Big(\begin{matrix}_0&_0\cr^0&^{\beta^{-1}\sigma}\end{matrix}\Big)\,,
\qquad
\Pi\,=\,\Big(\begin{matrix}_0&_1\cr^{-1}&^0\end{matrix}\Big)\,,\qquad H_t\,=\,
\frac{_1}{^2}\, p\cdot m^{-1}p\,+\,V_t(q)\,,&\cr
&F_t\,=\,(0,\,(\sigma-\gamma)m^{-1}p+f_t(q)),,\qquad
\zeta_t=(0,\,\xi_t)\,.
\nonumber
\qqq
The subtracted mean local velocity in the phase space has here the form
\qq
v_t\ =\ \big(m^{-1}p+\beta^{-1}\nabla_p\ln{\rho_t}\,,\ -\nabla V_t
-\gamma\,m^{-1}p+f_t-\beta^{-1}\nabla_q\ln{\rho_t}-\beta^{-1}
\sigma\,\nabla_p\ln{\rho_t}\big)
\qqq
and it vanishes for the Gibbs density $\ \rho(q,p)
=Z^{-1}\ee^{-\beta\,H(q,p)}\ $ in the equilibrium case where 
$\,\sigma=\gamma$, \,the potential 
$\,V_t\,$ is time-independent, and the non-conservative force $f_t\,$ 
is absent.

\subsubsection{Harmonic chain}

An example of a Langevin-Kramers dynamics is provided 
by a Fermi-Pasta-Ulam chain \cite{FPU} with ends coupled to a friction 
force and a white noise. Such chains were often used in the theoretical 
studies of the Fourier law \cite{BLR-B}. \,Here $\,q=(r^a_i)\,$ with 
$\,i=1,\dots,N$, $a=1,\dots,d$, \,and
\qq
&\gamma^{ab}_{ij}\,=\,\gamma_0\,\delta^{ab}
(\delta_{i1}\delta_{1j}+\delta_{iN}\delta_{Nj})\,,\qquad
\sigma^{ab}_{ij}\,=\,\gamma_0\,\delta^{ab}\big((1+\eta)\delta_{i1}
\delta_{1j}+(1-\eta)\delta_{iN}\delta_{Nj}\big)\,,&\cr
&m^{ab}_{ij}\,=\,m_0\,\delta^{ab}\,\delta_{ij}\,,\qquad
V(q)\,=\,\sum\limits_{i<N}^N U(r_{i\,(i+1)})\,+\,
\sum\limits_{i}U_0(\bm r_i)\,,&
\qqq
The dynamics in the bulk (i.e. for $\,i\not=1,N$) \,is purely Hamiltonian,
whereas the boundary degrees of freedom $\,\bm r_0\,$ and $\,\bm r_N\,$
are exposed to the thermal noise at temperatures $\,\beta^{-1}(1\pm\eta)$,
\,respectively, and to friction. \,The harmonic case (that does not lead 
to the Fourier law \cite{RLL}) with $\,U(r)=\frac{\kappa}{2}r^2\,$ and 
$\,U_0(\bm r)=\frac{k}{2}r^2\,$ corresponds to the linear stochastic
equation of the type (\ref{linear}) with the matrices
\qq
&M^{ab}_{ij}\ =\ \delta^{ab}\,\Big(
\begin{matrix}_0&_{m_0^{-1}\delta_{ij}}\cr
{}^{-(-\kappa\Delta+k)_{ij}}&^{-\gamma_0m_0^{-1}(\delta_{i1}\delta_{1j}
+\delta_{iN}\delta_{Nj})}\end{matrix}\Big)\,,&\cr
&D^{ab}_{ij}\ =\ \beta^{-1}\gamma_0\,\delta^{ab}\,\Big(\begin{matrix}_0&_0\cr
{}^0&^{(1+\eta)\delta_{i1}\delta_{1j}+(1-\eta)\delta_{iN}\delta_{Nj}}
\end{matrix}\Big)\,.&
\qqq
The covariance matrix of the invariant Gaussian measure has the form
\qq
C^{ab}_{ij}\ =\ \beta^{-1}\delta^{ab}\,
\Big(\begin{matrix}_{\big(\frac{1}{-\kappa\Delta+k}
\big)_{ij}}&_0\cr{}^0&^{m_0\delta_{ij}}\end{matrix}\Big)+\beta^{-1}\eta\,
\delta^{ab}\,\Big(\begin{matrix}_{X_{ij}}&_{Z_{ij}}\cr{}^{-Z_{ij}}&^{Y_{ij}}
\end{matrix}\Big)
\qqq
with matrices $\,X, Y, Z\,$ that may be calculated exactly \cite{RLL} 
\,(for $\,\eta=0$, \,it reduces to the covariance of the Gibbs measure). 
\,The subtracted mean local velocity is
\qq
v(q,p)\ =\ \big(M+DC^{-1}-\beta^{-1}\Pi C^{-1}\big)\,
\Big(\begin{matrix}{}q\cr{}p
\end{matrix}\Big),
\qqq
where the matrix on the right hand side has, up to terms quadratic in 
the relative temperature difference $\,\eta$, \,the entries
\qq
\eta\,m_0^{-1}\delta^{ab}\,\Big(\begin{matrix}
_{-\big(Z(-\kappa\Delta+k)\big)_{ij}}&_{m_0^{-1}Y_{ij}}\cr
{}^{-\big(Y(-\kappa\Delta+k)\big)_{ij}}&^{-\big((-\kappa\Delta+k)Z\big)_{ij}
-\gamma_0m_0^{-1}(\delta_{i1}Y_{1j}+\delta_{iN}Y_{Nj})+\gamma_0
(\delta_{i1}\delta_{1j}-\delta_{iN}\delta_{Nj})}\end{matrix}\Big).
\qqq
The Lagrangian flow of $\,v\,$ is obtained by the linear action of the 
matrix $\ \ee^{(M+DC^{-1}-\beta^{-1}\Pi C^{-1})(t-t_0)}\ $ which is 
straightforward to calculate in the linear order in $\,\eta$.

\nsection{Fluctuation-dissipation relations}
\label{sec:FDT}
\subsection{Equilibrium Fluctuation-Dissipation Theorem}

The equilibrium Fluctuation-Dissipation Theorem \cite{Nyq,CalWel,Kubo} relates
the spontaneous dynamical fluctuations in an equilibrium state to
the relaxation dynamics after a tiny perturbation out of the equilibrium. 
It holds for a wide class of equilibrium systems including the ones 
described by the equilibrium Langevin equation
\qq
\dot{x}_t^i&=&-\,\beta\,d^{ij}_t(x_t)\,(\partial_jH)(x_t)\,
+\,\pi^{ij}_t(x_t)\,(\partial_jH)(x_t)\,
-\,\beta^{-1}\,(\partial_j\pi^{ij}_t)(x_t)\,+\,r^i_t(x_t)\,
\,+\,\zeta^i_t(x_t)
\label{equileq2}
\qqq
of the type discussed above. We assume that the process $\,x_t\,$ has 
the time-independent Gibbs 
instantaneous PDF $\,\rho(x)=Z^{-1}\ee^{-\beta\,H(x)}\,$ and denote
by $\,\big\langle\,-\,\big\rangle\,$ the dynamical expectation. 
\,The FDT asserts that \cite{MPRV}
\qq
\partial_s\hspace{0.025cm}\big\langle\,O^1(x_s)\,O^2(x_t)\,\big\rangle\ 
=\ \beta^{-1}
\frac{_\delta}{^{\delta h_s}}\big|_{h=0}\,\big\langle\,O^2(x_t)\,\big\rangle_h
\label{FDT}
\qqq
for $\,s<t$, \,where $\,O^a(x)\,$ are functions (well behaved at infinity),
that we shall call (single-time) observables, and where on the right hand side 
the expectation $\,\big\langle\,-\,\big\rangle_h\,$ involves the process
obtained by replacing the Hamiltonian $\,H(x)\,$ in
the original dynamics (\ref{equileq2}) by its time-dependent perturbation 
$\,H(x)-h_t\,O^1(x)\,$ within some time interval. \,The left hand
side is the time derivative of the 2-time correlation function in
the dynamics determined by Eq.\,(\ref{equileq2}) and the right hand side
is the response of the single-time correlation function to a small
dynamical perturbation of the Hamiltonian of the system. The temperature
$\,\beta^{-1}\,$ appears as the coefficient relating the two functions.
For the sake of completeness, we give a proof of the FDT (\ref{FDT}) 
in Appendix D. \,It is often more convenient to consider the 
time-integrated version of the FDT:
\qq
\big\langle\,O^1(x_t)\,O^2(x_t)\,\rangle\,-\, 
\big\langle\,O^1(x_s)\,O^2(x_t)\,\rangle\ 
=\ \beta^{-1}
\frac{_\partial}{^{\partial h_0}}\big|_{h_0=0}\,\big\langle\,O^2(x_t)\,
\big\rangle_{h_0,s}\,,
\label{IFDT}
\qqq
where $\,\big\langle\,-\,\big\rangle_{h_0,s}\,$ corresponds to the
expectation where the original Hamiltonian $\,H(x)\,$ is replaced
starting at time $\,s<t\,$ by its time-independent perturbation 
$\,H(x)-h_0O^1(x)$. 

\subsection{Modified Fluctuation Dissipation Theorem}

We may immediately apply the FDT to the Lagrangian-frame process
$\,\tilde x_t\,$ obtained from the process $\,x_t\,$ satisfying 
the Langevin equation (\ref{nonequileq1}). \,Indeed,
as was shown in Sect.\,\ref{sec:modcurr}, the process $\,\tilde x_t
=\Phi_t^{-1}(x_t)$, \,where $\,\Phi_t\,$ is the flow of the
subtracted mean local velocity (\ref{locvels}), \,satisfies the 
equilibrium stochastic
equation (\ref{equileq1}) and has the time-independent instantaneous
PDF $\,\rho_{t_0}(\tilde x)=Z^{-1}\ee^{-\beta\,\tilde H(\tilde x)}$.
\,We infer that for observables $\,\tilde O^a(\tilde x)$,
\qq
\partial_s\hspace{0.025cm}\big\langle\,\tilde O^1(\tilde x_s)
\,\tilde O^2(\tilde x_t)
\,\rangle\ =\ \beta^{-1}\frac{_\delta}{^{\delta\tilde h_s}}
\big|_{\tilde h=0}\,\big\langle\,\tilde O^2(\tilde x_t)\,
\big\rangle_{\tilde h}
\label{LFDT}
\qqq
where $\,\big\langle\,-\,\big\rangle_{\tilde h}\,$ involves the process 
obtained by replacing the Hamiltonian 
$\,\tilde H(\tilde x)\,$ in the Lagrangian-frame dynamics (\ref{equileq1}) 
by its time-dependent perturbation $\,\tilde H(\tilde x)-\tilde 
h_t\tilde O^1(\tilde x)$ during a time interval. \ Observe
that this perturbation corresponds to the replacement of the
Hamiltonian $\,H_t(x)\,$ in the original equation (\ref{nonequileq1})
for $\,x_t\,$ by $\,H_t(x)-\tilde h_t\tilde O^1(\Phi_t^{-1}(x))$. \,Indeed, 
the latter replacement adds the term
\qq
\beta\,\tilde h_t\,b^{ij}_t(x_t)\,\partial_{x^j}
\tilde O^1(\Phi_t^{-1}(x))|_{x=x_t}
\label{6.5}
\qqq 
on the right hand side of Eq.\,(\ref{nonequileq1}) and, in virtue
of Eq.\,(\ref{tosubs}), results in the additional term 
\qq
&&(\partial_k\Phi_t^{-1})^i(x_t)\,\Big[\beta\,\tilde h_t\,b^{kl}_t(x_t)\,
\partial_{x^l}|_{x=x_t}\tilde O^1(\Phi_t^{-1}(x))\Big]\cr\cr
&&=\ \beta\,\tilde h_t\,(\partial_k\Phi_t^{-1})^i(x_t)\,b^{kl}_t(x_t)\,
(\partial_l\Phi_t^{-1})^j(x_t)\,(\partial_j\tilde O^1)(\tilde x_t)\cr\cr
&&=\ \beta\,\tilde h_t\,\tilde b^{ij}_t(\tilde x_t)\,
(\partial_j\tilde O^1)(\tilde x_t)
\label{6.6}
\qqq
with $\ \tilde b^{ij}_t=\tilde d^{ij}_t-\beta^{-1}\tilde\Pi^{ij}_t\,$
in Eq.\,(\ref{equileq1}) for $\,\tilde x_t=\Phi_t^{-1}(x_t)\,$
(with the same transformations $\,\Phi_t\,$ as in the unperturbed process). 
\,Upon defining the Eulerian-frame time-dependent observables 
\qq
O^a_t(x)\ =\ \tilde O^a(\Phi_t^{-1}(x))\,,
\label{OtO}
\qqq
the Lagrangian-frame FDT (\ref{LFDT}) may be rewritten as the identity
\qq
\partial_s\hspace{0.025cm}\big\langle\,O^1_s(x_s)\,O^2_t(x_t)\,\rangle\ 
=\ \beta^{-1}
\frac{_\delta}{^{\delta\tilde h_s}}\big|_{\tilde h=0}\,\big\langle\,O^2_t(x_t)\,
\big\rangle_{\tilde h}\,.
\label{LFDT'}
\qqq
Note that the time-dependent observables $\,O^a_t(x)\,$ are constant 
along the Lagrangian trajectories
of the velocity (\ref{locvels}): $\,O^a_t(\Phi_t(x))=\tilde O^a(x)$.
\,In other words, they obey the scalar advection equation  
\qq
\partial_tO^a_t\,+\,v_t\cdot\nabla O^a_t\,=\,0\,
\label{sadveq}
\qqq
and are frozen in the Lagrangian frame of the subtracted mean
local velocity $\,v_t$. \ Since the values of the time-dependent observable
$\,O^1\,$ may be chosen arbitrarily at time $\,s\,$ and that of
$\,O^2\,$ at time $\,t$, \,the only trace of time dependence of 
the observables $\,O^a\,$ in the identity (\ref{LFDT'}) for fixed 
pair of times $\,s<t\,$ enters through the time derivative 
$\,\partial_s\,$ on the left hand side that differentiates also 
the explicit time-dependence of $\,O^1\,$ determined
by Eq.\,(\ref{sadveq}). \,We may then rewrite Eq.\,(\ref{LFDT'}) using 
observables frozen in the Eulerian frame as the Modified 
Fluctuation-Dissipation Theorem, 
\qq
\partial_s\hspace{0.025cm}\big\langle\,O^1(x_s)\,O^2(x_t)\,\big\rangle\ -\ 
\big\langle\,(v_s\cdot\nabla O^1)(x_s)\,O^2(x_t)\,\big\rangle
=\ \beta^{-1}
\frac{_\delta}{^{\delta h_s}}\big|_{h=0}\,\big\langle\,O^2_t(x_t)\,
\big\rangle_{h}\,,
\label{MFDT}
\qqq  
where the expectation $\,\big\langle\,-\,\big\rangle_h\,$ on the right
hand side refers now to the process obtained by replacing 
the Hamiltonian $\,H_t\,$ in Eq.\,(\ref{nonequileq1})    
by $\,H_t(x)-h_t\,O^1(x)$. \,In the time-integrated form, Eq.\,(\ref{MFDT})
becomes
\qq
\big\langle\,O^1(x_t)\,O^2(x_t)\,\rangle\,-\, 
\big\langle\,O^1(x_s)\,O^2(x_t)\,\rangle\ -\ 
\int\limits_s^t\big\langle\,(v_\sigma\cdot\nabla O^1)(x_\sigma)
\,O^2(x_t)\,\big\rangle\,d\sigma\,\,\cr\cr
=\ \beta^{-1}
\frac{_\partial}{^{\partial h}}\big|_{h_0=0}\,\big\langle\,O^2(x_t)
\,\big\rangle_{h_0,s}\,,
\label{IMFDT}
\qqq
with a corrective integral term with respect to Eq.\,(\ref{IFDT}).
\,An experimental check of the time-integrated MFDT for a
colloidal particle has been described in \cite{GPCCG}.
Fig.\,4 shows the numerical check of this relation and of its 
Lagrangian-frame counterpart for the stationary process solving the 
SDE (\ref{1dSDE}) for $\,O^a(\theta)=sin(\theta)$. 
\vskip 0.2cm

\begin{figure}[h!]
\label{fig:fig4}
\begin{center}
\vskip -0.2cm
\leavevmode
{%
\hspace*{0.6cm}
\begin{minipage}{0.4\textwidth}
        \includegraphics[width=5.5cm,height=7.2cm,angle=-90]{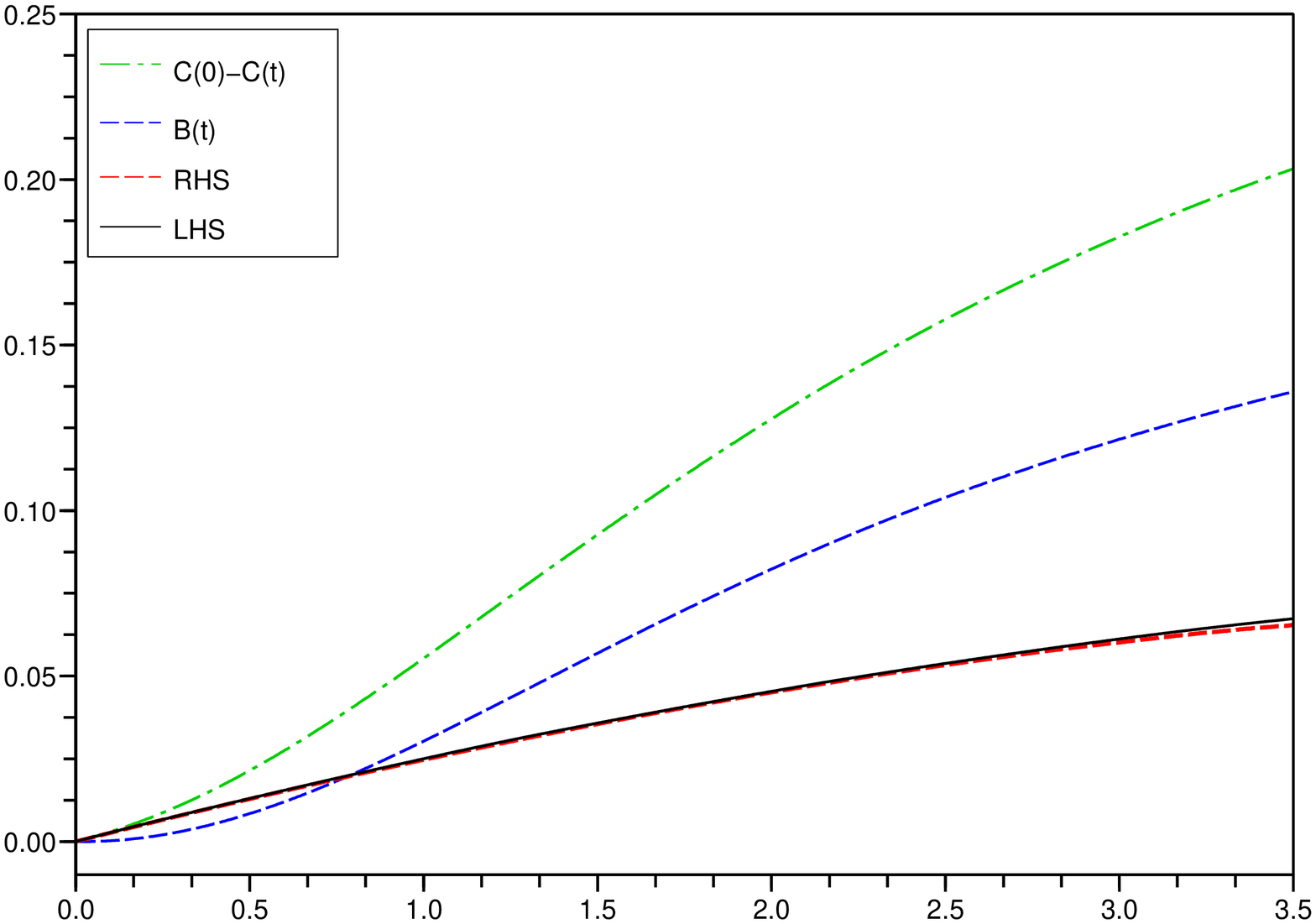}\\
        \vspace{-0.8cm} \strut
        \end{minipage}}
    \hspace*{0.6cm}
{%
      \begin{minipage}{0.4\textwidth}
        \includegraphics[width=5.5cm,height=7.2cm,angle=-90]{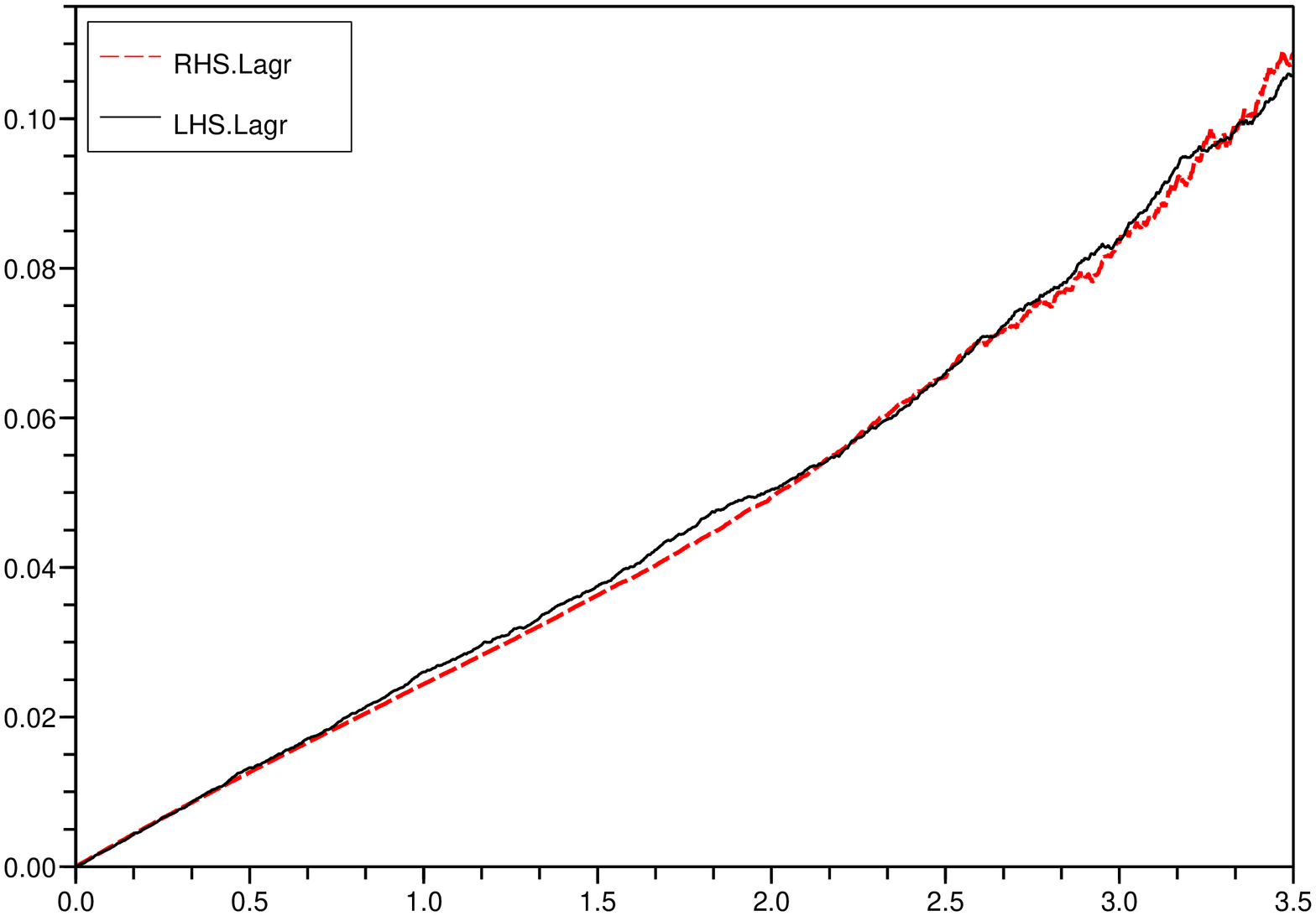}\\
      \vspace{-0.8cm} \strut
        \end{minipage}}\hspace*{0.5cm}
\hspace*{0.7cm}
\end{center}
\hspace*{1.4cm}\parbox{13.6cm}{
\caption{\small{\,Left: \ \ \ the bottom coinciding curves: LHS (continuous black) and RHS 
(dashed red)\break\hspace*{2.7cm}of the integrated MFDT (\ref{IMFDT}) for
$\,O^a(\theta)=\sin(\theta)$, the upper 
(dot-dashed\break\hspace*{2.73cm}green) curve: the first two terms 
on its LHS, the middle (dashed blue)\break\hspace*{2.73cm}curve: the 
corrective integral term\hfill\break\hspace*{1.58cm}\vspace{-0.3cm} 
\break\hspace*{1.58cm}Right: \hspace{0.1cm}RHS (dashed red curve) 
and LHS (continuous black curve) of the
integrated-\break\hspace*{2.71cm}in-time Lagrangian-frame FDT (\ref{LFDT}) 
\,with $\,\tilde O^a(\tilde\theta)=\sin(\tilde\theta)$\hfill}} }
\end{figure}
\vskip 0.2cm

The MFDT was proven directly in \cite{CFG} in the stationary setup 
and shown to be equivalent to identity (\ref{LFDT'}) similar to the
equilibrium FDT (\ref{FDT}) but for observables frozen in the Lagrangian
frame of mean local velocity. In the present paper, we unravel
the deeper reason for that equivalence, namely the fact that 
the non-equilibrium diffusion process observed in the Lagrangian frame
of the (subtracted) mean local velocity evolves according to an equilibrium 
dynamics with a time-independent instantaneous PDF.

\subsection{Links with fluctuation relations}

Ref.\,\cite{CFG}) also 
discussed fluctuation relation extending the MFDT to non-stationary
situations. It was shown there that the Hatano-Sasa version \cite{HatSas} 
of the Jarzynski equality \cite{Jarz1,Jarz2} reduces close to stationarity
to the MFDT for special observables and that one need Croooks' extention 
\cite{Crooks2} of the Jarzynski-Hatano-Sasa equality to extract at 
stationarity the MFDT for general observables. The results of the present 
paper permit to propose yet another extension of the MFDT. 
\vskip 0.1cm

For the process $\,x_t\,$ evolving accordingly to the Langevin equation
(\ref{equileq}) but with a time-dependent Hamiltonian $\,H_t(x)$,
\,the Jarzynski equality reads
\qq
\big\langle\,\ee^{-\beta\,W_{t_0,t}}\,\big\rangle\ =\ \frac{{Z_t}}{{Z_{t_0}}}\,,
\label{jarzeq} 
\qqq
\vskip -0.4cm
\noindent where 
\vskip -0.6cm
\qq
W_{t_0,t}\ =\ \int\limits_{t_0}^t(\partial_sH)(x_s)\,ds\,\qquad{\rm and}\qquad
Z_t =\ \int\ee^{-\beta\,H_t(x)}\,dx\,, 
\qqq
provided that the PDF of $\,x_{t_0}\,$ is $\,Z_{t_0}^{-1}
\ee^{-\beta H_{t_0}}$. \,Applied to case with the Hamiltonian 
$\,H_t(x)=H(x)-\hspace{-0.07cm}\sum\limits_{a=1,2}
\hspace{-0.07cm}h^a_t\,O^a(x)\,$ and expanded to the second order
in functions $\,h^a_t$, \ Eq.\,(\ref{jarzeq}) reduces to the FDT (\ref{FDT}).
The proof goes as in \cite{CG1} where it was written for a less general case. 
\vskip 0.1cm

The above observations apply to the case of the Lagrangian-frame dynamics.
For the process $\,\tilde x_t\,$ satisfying 
the SDE (\ref{equileq1}) but with the Hamiltonian $\,\tilde H\,$
replaced by $\,\tilde H(\tilde x)+\tilde H'_t(\tilde x)\,$ with
$\,\tilde H'_{t}=0\,$ for $\,t\leq t_0$, \,we have for $\,t>t_0\,$
the Lagrangian-frame version of the Jarzynski equality:
\qq
\big\langle\,\ee^{-\beta\,\tilde W_{t_0,t}}\,\big\rangle\ =\ 
\frac{\tilde Z_t}{\tilde Z}\qquad{\rm for}\qquad
\tilde W_{t_0,t}\ =\ \int\limits_{t_0}^t(\partial_s\tilde H'_s)(\tilde x_s)\,ds
\label{LJarz}
\qqq
and $\ \tilde Z_t=\int\limits\ee^{-\beta\,(\tilde H+\tilde H'_t)
(\tilde x)}d\tilde x$, \ provided that the PDF
of $\,\tilde x_{t_0}\,$ is $\,\tilde Z^{-1}\ee^{-\beta\,\tilde H}
=\rho_{t_0}$. \,The process $\,x_t\,$ such that 
$\,\tilde x_t=\Phi_t^{-1}(x_t)$, \,with $\,\Phi_t\,$ standing for the Lagrangian
flow of the mean local velocity $\,v_t(x)\,$ of the unperturbed process
$\,x_t$, \,satisfies the SDE (\ref{nonequileq1}) with the original Hamiltonian
$\,H_t(x)\,$ replaced by $\,H_t(x)+H'_t(\Phi_t^{-1}(x))$. \,This follows
by the same argument as around Eqs.\,(\ref{6.5}) and (\ref{6.6}).
\,In terms of the perturbed process $\,x_t$,
\qq
\tilde W_{t_0,t}\ =\ \int\limits_{t_0}^t(\partial_s\tilde H'_s)
\big(\Phi_t^{-1}(x_s)\big)\,ds\qquad{\rm and}\qquad
\frac{\tilde Z_t}{\tilde Z}\ =\ \int\ee^{-\beta\,H'_t(x)}
\rho_{t_0}(x)\,dx\,.    
\qqq
For $\ \tilde H'(\tilde x)=-\hspace{-0.07cm}\sum\limits_{a=1,2}h^a_t
\,\tilde O^a(\tilde x)$, \ one obtains the Lagrangian-frame FDT
(\ref{LFDT}) equivalent to the MFDT (\ref{MFDT}) by expanding
the identity (\ref{LJarz}) to the 2$^{\rm nd}$ order in $\,h^a_t$.
\,Not very surprisingly, there exist different fluctuation relations
that may be viewed as an extension of the MFDT to more general 
situations.

\nsection{Non-equilibrium diffusions without Lagrangian picture} 
\label{sec:obstr}

In the preceding sections, we have discussed diffusion processes in a finite 
dimensional phase space. The basic assumption underlying the discussion 
of the Lagrangian-frame picture of diffusions was the existence of 
the Lagrangian flow $\,x\mapsto\Phi_t(x)\,$ of the mean local velocity 
satisfying Eqs.\,(\ref{Phieq}). This is guaranteed if the velocity 
$\,v_t(x)\,$ is smooth and the (phase-)space $\,\CX\,$ is compact, 
like in the circle example, but may be not assured if $\,\CX\,$ is 
unbounded in which case the Lagrangian trajectories of $\,v_t\,$ 
may blow up in finite time. The idea of the decoupling of probability 
flux by the passage to the Lagrangian frame of the mean local velocity 
can, in principal, be applied to infinite-dimensional diffusive processes. 
It appears, however, that a number of known examples of diffusive 
processes described by stochastic PDEs do not allow a global flow 
of mean local velocity and, hence, do not admit a Lagrangian-frame 
equilibrium-like description. Let us illustrate this phenomenon 
in specific cases.

\subsection{One-dimensional Kardar-Parisi-Zhang equation}

The KPZ stochastic PDE \cite{KPZ} describes the fluctuations of a 
$\,d$-dimensional interface with the height function $\,h_t(x)$. 
\,It has the form
\qq
\partial_t h_t(x)\ =\ \nu\nabla^2h_t(x)\,
+\,\frac{_1}{^2}\lambda(\nabla h_t(x))^2\,+\,\zeta_t(x)
\qqq
where $\,\zeta_t(x)\,$ is the white noise with the covariance
\qq
\big\langle\,\zeta_t(x)\ \zeta_s(y)\,\big\rangle\ =\ 
2\,D\,\delta(t-s)\,\delta(x-y)\,.
\label{covKPZ}
\qqq
The adjoint generator of the process $\,h_t\,$ in the  (infinite-dimensional)
space of the height functions $\,h\,$ has the form
\qq
L^\dagger\ =\ \int\frac{\delta}{\delta h(x)}\Big[-\,\nu\nabla^2h(x)
\,-\,\frac{_\lambda}{^2}(\nabla h(x))^2\,+\,D\frac{\delta}{\delta h(x)}
\,\Big]dx\,.
\qqq
A straightforward (although somewhat formal) calculation \cite{HalZhang} 
shows that in one space-dimension with periodic boundary conditions (where
$\,\nabla h=\partial_xh$), \,the Gaussian density in the space of height 
functions
\qq
\rho[h]\ =\ Z^{-1}\ee^{-\frac{\nu}{2D}\int(\nabla h(x))^2dx}
\qqq
is annihilated by $\,L^\dagger\,$ (for all values of $\,\lambda$) \,and thus 
stays invariant. \,The corresponding  mean local velocity given 
by Eq.\,(\ref{mlv}) has the form
\qq
v[h](x)\ =\ \frac{_1}{^2}\lambda(\nabla h_t(x))^2
\qqq
and the Lagrangian trajectories of $\,v[h]\,$ should be solutions
of the equation
\qq
\partial_th_t\,=\,\frac{_1}{^2}\lambda(\nabla h_t(x))^2
\label{Lagrh}
\qqq
that becomes for $\,u_t(x)=-\lambda\nabla h_t(x)\,$ the inviscid
Burgers equation \cite{Burgers}
\qq
\partial_tu_t(x)\,+\,u_t(x)\nabla u_t(x)\,=\,0
\qqq
with the solutions satisfying the relation 
\qq
u_t(x+(t-t_0)u_{t_0}(x))=u_{t_0}(x)
\qqq
and developing discontinuities (shocks) for the first time $\,t_s>t_0\,$ 
such that $\,t_s=t_0+\frac{x_2-x_1}{u_{t_0}(x_2)-u_{t_0}(x_1)}\,$ for 
a pair of points 
$\,(x_1,x_2)$. \,The corresponding height function $\,h_t(x)$ looses at 
$\,t=t_s\,$ the differentiability and, although weak solutions of the 
inviscid Burgers equation exist beyond the time $\,t_s$, there is no unique 
global invertible Lagrangian flow of the mean local velocity $\,v[h]\,$ and no
global Lagrangian-frame picture of the KPZ evolution.

\subsection{Diffusive hydrodynamical limits}


Similar problems obstruct the existence of the Lagrangian
picture in the effective equations describing the 
large-deviations regime of fluctuations around diffusive 
hydrodynamical limits of some lattice particle systems. The evolution 
of the particles consists of random jumps to nearby sites. On the 
scales of the order of the size of the system $\,L$, \,and for times 
of the order $\,L^2$, \,such stochastic evolution gives 
rise to an effective diffusion in the space of macroscopic densities 
$\,n_t(x)$ \cite{Spohn,KipLan}. The dynamics of the densities 
is given by the continuity equation $\ \partial_tn_t+\nabla\cdot j_t=0\ $
for 
\qq
j^i_t(x)\ =\ -\frac{_1}{^2}D^{ij}(n_t(x))\,\partial_jn_t(x)\,
+\,\zeta^i_t(x|n_t)
\qqq
where $\,\zeta_t(x|n)\,$ is the density-dependent white noise in time 
and space with the covariance
\qq
\big\langle\,\zeta^i_t(x|n)\ \zeta^j_s(y|n)\,\big\rangle\ =\ 
\epsilon\,\delta(t-s)\,\delta(x-y)\,\chi^{ij}(n(x))\,,
\label{coveta}
\qqq
where $\,\epsilon^{-1}\propto L^{-d}\,$ is the total number of microscopic 
particles assumed to be large. In particular, in the limit where 
$\,\epsilon=0$, \,the density $\,n_t(x)\,$ satisfies the
deterministic hydrodynamical-limit diffusion equation
\qq
\partial_tn_t\,=\,\frac{_1}{^2}\partial_i\big(D^{ij}(n_t(x))\,
\partial_jn_t(x)\big)\,.
\label{hydlim}
\qqq
One considers such systems with periodic 
boundary conditions or with Dirichlet ones where one fixes the boundary 
values of the density $\,n_t(x)\,$ on the boundary of a finite domain 
$\,\Lambda\subset\NR^d$.
\,The first case corresponds to an equilibrium evolution whereas the 
second one to a non-equilibrium boundary-driven one. The adjoint
generator of the process $\,n_t\,$ has the form
\qq
L^\dagger\ =\ \frac{_1}{^2}\int\Big(\partial_i
\frac{\delta}{\delta n(x)}\Big)\Big[
D^{ij}(n(x))\,\partial_jn(x)\,+\,\epsilon\,\chi^{ij}(n(x))\,\partial_j
\frac{\delta}{\delta n(x)}\Big]dx
\qqq
up to the terms of higher orders in $\,\epsilon$.
To the leading order, the stationary PDF in the space
of density functions takes the semi-classical form
\qq
\rho[n]\ =\ \ee^{-\frac{1}{\epsilon}S[n]}
\qqq
with the functional $\,S[n]\,$ satisfying the Hamilton-Jacobi equation
\qq
\int\Big(\partial_i\frac{\delta S}{\partial n(x)}\Big)\Big[\chi^{ij}(n(x))\,
\partial_j\frac{\delta S}{\delta n(x)}\,-\,D^{ij}(n(x))
\,\partial_jn(x)\Big]dx\ =\ 0
\qqq
and a certain stability condition \cite{Bertini}. 
\,According to Eq.\,(\ref{mlv}), the mean
local velocity in the space of densities has the form
\qq
v[n](x)\ =\ \frac{_1}{^2}\Big[\partial_i\big(D^{ij}(n_t(x))\,
\partial_jn(x)\big)\,-\,\partial_i\Big(\chi^{ij}(n(x))\,\partial_j
\frac{\delta S}{\delta n(x)}\Big)\Big]
\label{mlvhl}
\qqq
up to terms that vanish at $\,\epsilon=0$.
\,The functional $\,S[n]\,$ is explicitly known in few 
boundary driven non-equilibrium situations for which one may
study the existence of the Lagrangian trajectories of $\,v[h]$.

\subsubsection{Zero range processes}

\noindent Here, $\,D^{ij}(n(x))=\varphi'(n(x))\,\delta^{ij}\,$ and
$\,\chi^{ij}(n(x))=\varphi(n(x))\,\delta^{ij}\,$ for an increasing
function $\,\varphi\geq0\,$ of $\,n\geq 0\,$ related explicitly to the
jump rates of the zero-range particle dynamics \cite{KipLan}.
The hydrodynamical-limit equation (\ref{hydlim}) reduces to the
form
\qq
\partial_tn_t(x)\ =\ \frac{_1}{^2}\nabla^2\varphi(n(x))\,.
\label{hydlim0r}
\qqq
and the functional $\,S[n]\,$ satisfies the relation \cite{Bertini}
\qq
\frac{\delta S}{\delta n(x)}\ =\ \ln\varphi(n(x))\,-\,\ln\lambda(x)\,,
\label{late}
\qqq
where $\,\lambda(x)=\varphi(\bar n(x))$, \,with $\,\bar n(x)\,$
providing the stationary solution of Eq.\,(\ref{hydlim0r}) so that
$\,\lambda(x)\,$ is a harmonic function on the domain $\,\Lambda\,$ with 
prescribed boundary values. \,In virtue of Eq.\,(\ref{late}),
\qq
&&\partial_i\Big(\chi^{ij}(n(x))\,\partial_j
\frac{\delta S}{\delta n(x)}\Big)\ =\ \nabla\cdot\varphi(n(x))\nabla\,
\big[\ln\varphi(n(x))-\ln\lambda(x)\big]\cr
&&=\ \nabla^2\varphi(n(x))\,-\,\nabla\cdot\big(\varphi(n(x))
\,\nabla\ln\lambda(x)\big)\,.
\qqq
One infers that in this case
\qq
v[n](x)\ =\ \frac{_1}{^2}\nabla\cdot\big(\varphi(n(x))\,
\nabla\ln\lambda(x)\big)\,.
\qqq
The equation for the Lagrangian trajectories of $\,v[n]\,$ has the form
\qq
\partial_tn_t(x)\ =\ \frac{_1}{^2}\varphi'(n(x))\hspace{0.03cm}
\big(\nabla n(x)\big)\cdot
\frac{\nabla\lambda(x)}{\lambda(x)}\,-\,\frac{_1}{^2}\varphi(n(x))\,
\frac{(\nabla\lambda)^2(x)}{\lambda^2(x)}
\qqq 
which is a quasi-linear $1^{\rm st}$-order PDE whose solutions may be
composed from characteristic curves.  The existence 
of global solutions will again be obstructed by caustics, i.e. by 
crossings of the projection of the characteristics to the space. That 
this phenomenon takes really place may be easily seen in one dimension 
where $\,\lambda(x)\,$ is a linear function.

\subsubsection{Symmetric simple exclusion process (SSEP)}

Here $\,D^{ij}=\delta^{ij}\,$ and $\,\chi^{ij}(n)=n(1-n)$. 
\,The functional $\,S[n]\,$ is explicitly known in one space-dimension 
\cite{DLS}. It satisfies the identity \cite{Bertini}
\qq
\frac{\delta S}{\delta n(x)}\ =\ \ln\frac{n(x)}{1-n(x)}\,-\,\varphi(x|n)
\qqq
where $\,\varphi(x|n)\,$ is the solution of the ordinary differential
equation
\qq
\frac{\nabla^2\varphi(x)}{(\nabla\varphi)^2(x)}\,+\,
\frac{1}{1+\ee^{\varphi(x)}}\ =\ n(x)
\qqq
with prescribed boundary values. The mean local velocity has the form
\qq
v[n](x)\ =\ \frac{_1}{^2}\nabla\big(n(x)(1-n(x))\nabla\varphi(x|n)\big)\,.
\qqq
We do not know if there are obstructions to the existence
of the corresponding Lagrangian flow.

\nsection{Conclusions}
\label{sec:concl}

We have shown that non-equilibrium Markov diffusions become
equilibrium ones when viewed in the Lagrangian frame of their mean 
local velocity. More exactly, the diffusion process transformed
to that frame, although in general non-stationary, satisfies the 
detailed balance and has instantaneous probability density that 
does not change in time and is equal to the Eulerian invariant 
density if the original process is stationary. The passage to 
the Lagrangian frame 
decouples the non-zero probability current from the non-equilibrium 
process. The equilibrium nature of the Langevin-frame process
explains on a deeper level the equilibrium-like fluctuation-dissipation 
relations observed in the Lagrangian-frame of mean local velocity 
in \cite{SpS0,CFG}. Our analysis indicates that the equilibrium and 
non-equilibrium diffusions are closer than usually perceived and 
the entire difference between them may be encoded in the probability 
current that does not vanish in the non-equilibrium case. This seems 
to be an interesting observation on the fundamental level. In practice, 
although the passage to the Lagrangian frame may be realized numerically 
in simulations of small systems, its experimental realization is far 
from obvious and its use in the analysis of stationary non-equilibrium 
dynamics may be hampered by the absence of knowledge of the invariant 
measure that enters the expression for the mean local velocity. 
As we have also seen, our arguments apply only to diffusive systems 
with the global flow of mean local velocity. Such global flow
is absent in important examples of non-equilibrium diffusions 
described by stochastic partial differential equations.   
It remains to be seen to what extent a similar analysis may be 
carried through for other models of non-equilibrium dynamics.

\nappendix{A}

\noindent We check here the formula (\ref{prdel}) for the probability
current (\ref{curr}). First note that for a similar average
as in Eq.\,(\ref{prdel}) but with the right time derivative,
\qq 
&&\big\langle\,\dot{x}^i_{t+}\,\delta(x-x_t)\,\big\rangle\ \equiv\  
\lim\limits_{\epsilon\to0}\,\,\Big\langle\,
\frac{x^i_{t+\epsilon}-x^i_t}{\epsilon}
\,\,\delta(x-x_t)\,\Big\rangle\cr\cr
&&=\ \lim\limits_{\epsilon\to0}\,\,\frac{1}{\epsilon}\,\rho_t(x)\,\Big(\int
P(t,x;t+\epsilon,y)\,y^i\,dy\,-\,x^i\Big)\ =\ 
\rho_t(x)\,(L_tx^i)\cr\cr
&&=\ \big[\hat u^i_t(x)\,+\,(\partial_jd^{ij}_t)(x)\big]\,
\rho_t(x)\,.
\label{rhtder}
\qqq
On the other hand, for the left time derivative,
\qq
&&\big\langle\,\dot{x}^i_{t-}\,\delta(x-x_t)\,\big\rangle\ \equiv\  
\lim\limits_{\epsilon\to0}\,\,\Big\langle\,
\frac{x^i_{t}-x^i_{t-\epsilon}}{\epsilon}
\,\,\delta(x-x_t)\,\Big\rangle\cr\cr
&&=\ \lim\limits_{\epsilon\to0}\,\,\frac{1}{\epsilon}\,\Big(\rho_t(x)\,x^i
\,-\,\int\rho_{t-\epsilon}(y)\,y^i\,P(t-\epsilon,y;t,x)\,dy\Big)
\ =\ (L_t^\dagger\rho_t)(x)\,x^i\,-\,L_t^\dagger(\rho_t(x)\,x^i)\cr\cr
&&=\ [\hat u^i_t(x)\,-\,(\partial_j d^{ij}_t)(x)\,-\,
2\,d^{ij}_t(x)\,\partial_j\big]\,\rho_t(x)\,,
\label{lftder}
\qqq
where the second equality combined the derivatives over $\,\epsilon\,$ 
of $\,\rho_{t-\epsilon}\,$ and of $\,P(t-\epsilon,y;t,x)$. \,The addition
of the relations (\ref{lftder}) and (\ref{rhtder}) gives the
identity (\ref{prdel}).

\nappendix{B}

\noindent Let us check that under the change of variables 
$\ x\longmapsto x'=\Psi(x)$, \ the 
mean local velocity (\ref{mlv}) transforms as a vector field. \,In new 
variables, the process $\,x'_t=\Psi(x_t)\,$ satisfies the Stratonovich 
stochastic equation
\qq
\dot{x}'^i\,= u'^i_t(x')\,+\,\zeta'_t(x')
\qqq
\vskip -0.4cm
\noindent with 
\vskip -0.6cm
\qq
&&u'^i_t(x')\,=\,(\partial_k\Psi)^i(x)\,u^k_t(x)\,,\qquad
\zeta'^i_t(x')\,=\,(\partial_k\Psi)^i(x)\,\zeta^k_t(x)
\label{upru}
\qqq
for $\,x'=\Psi(x)$. \ 
The covariance of the white noise $\,\zeta'_t(x')\,$ is
\qq
\big\langle\,\zeta'^i_t(x')\ \zeta^j_s(y')\,\big\rangle\ =\ 2\,\delta(t-s)\,
D'^{ij}_t(x',y')
\qqq
\vskip -0.4cm
\noindent for 
\vskip -0.6cm
\qq 
D'^{ij}_t(x',y')\,=\,(\partial_k\Psi)^i(x)\,D^{kl}_t(x,y)\,
(\partial_l\Psi)^j(y)
\qqq
and $\,x'=\Psi(x),\ y'=\Psi(y)$. \,The instantaneous PDF of the process 
$\,x'_t\,$ is
\qq
\rho'_t(x')\ =\ \rho_t(x)\,\Big(\frac{\partial(\Psi(x))}{\partial(x)}
\Big)^{-1}\,,
\qqq
where $\,\frac{\partial(\Psi((x))}{\partial(x)}\,$ stands for the Jacobian
of the change of variables. \,In the new variables, the mean local velocity 
(\ref{mlv}) is
\qq
v'^i(x')\ =\ \hat u'^i_t(x')\,-\,d'^{ij}_t(x')\,
(\partial_{j}\ln{\rho'_t})(x')\,,
\qqq
\vskip -0.4cm
\noindent where
\vskip -0.6cm
\qq
d'^{ij}_{t}(x')\,=\,D'^{ij}_{t}(x',x')\qquad{\rm and}\qquad  
\hat u'^i_t(x')\,=\,u'^i_t(x')\,-\,r'^i_t(x')\,.
\label{standpr}
\qqq
The deterministic correction
\qq
r'^i_t(x')&=&\partial_{y'^j}D'^{ij}_{t}(x',y')|_{y'=x'}\ =\ 
(\partial_{j}\Psi^{-1})^h(y')\,\,\partial_{y^h}\Big[(\partial_k\Psi)^i(x)
\,D^{kl}_t(x,y)\,(\partial_l\Psi)^j(y)\Big]\Big|_{y=x}\cr\cr
&=&(\partial_k\Psi)^i(x)\,\partial_lD^{kl}_t(x,y)|_{y=x}\,+\,
(\partial_{j}\Psi^{-1})^h(x')\,
(\partial_k\Psi)^i(x)\,d^{kl}_t(x)\,(\partial_h\partial_l\Psi)^j(x)\cr\cr
&=&(\partial_k\Psi)^i(x)\,\Big[r^k_t(x)\,+\,
d^{kl}_t(x)\,(\partial_{j}\Psi^{-1})^h(x')\,
(\partial_h\partial_l\Psi)^j(x)\Big].
\label{a:r}
\qqq
On the other hand, using the standard formula for the derivative of the
logarithm of a determinant, we obtain
\qq
(\partial_l\Psi)^j(x)\,(\partial_{j}\ln{\rho'_t})(x')&=&
(\partial_l\Psi)^j(x)\,(\partial_{j}\Psi^{-1})^h(x')
\,\partial_{h}\Big[\ln{\rho_t(x)}-\ln{\frac{\partial(\Psi(x))}
{\partial(x)}}\Big]\cr\cr
&=&(\partial_l\ln{\rho_t})(x)\ -\ 
\,(\partial_{j}\Psi^{-1})^h(x')\,(\partial_l\partial_h\Psi)^j(x)\,.
\label{a:ln}
\qqq
Hence
\qq
r'^i_t(x')\,
+\,d'^{ij}_t(x')\,(\partial_{j}
\ln{\rho'_t})(x')
\ =\ (\partial_k\Psi)^i(x)\,\Big[r^k_t(x)\,+\,
d^{kl}_t(x)\,(\partial_{j}\Psi^{-1})^h(x')\,
(\partial_h\partial_l\Psi)^j(x)\Big]\cr\cr
+\ (\partial_k\Psi)^i(x)\,d^{kl}_t(x)\,\Big[(\partial_l\ln{\rho_t})(x)\ -\ 
\,(\partial_{j}\Psi^{-1})^h(x')\,(\partial_l\partial_h\Psi)^j(x)\Big]\cr\cr
=\ (\partial_k\Psi)^i(x)\,\Big[\hat r^k_t(x)\,
+\,d^{kl}_t(x)\,(\partial_l\ln{\rho_t})(x)\Big].\hspace*{1.92cm}
\label{B10}
\qqq
Finally, using also the 1$^{\rm st}$ of Eqs.\,(\ref{upru}), we obtain
the identity
\qq
v'^i_t(x')&=&u'^i_t(x')\,-\,r'^i_t(x')\,
-\,d'^{ij}_t(x')\,(\partial_{j}
\ln{\rho'_t})(x')\cr\cr
&=&(\partial_k\Psi)^i(x)\,\Big[u^k_t(x)\,-\,r^k_t(x)\,
-\,d^{kl}_t(x)\,(\partial_l
\ln{\rho_t})(x)\Big]\ =\ (\partial_k\Psi)^i(x)\,v^k_t(x)\,,
\qqq
which was to be shown.

\nappendix{C}

\noindent We give here the explicit formula for 
the time-dependent noise covariance $\,\tilde D_t\,$ of 
the Lagrangian-frame process corresponding to the harmonic 
Rouse polymer in linear shearing flow considered in Sect.\,\ref{sec:shear},
keeping the notations of that section.
$\,\tilde D_t\,$ is composed of $\,3\times3\,$ diagonal Fourier blocs
\qq
\tilde D^{ab}_{\ell,t}\ =\ (\gamma\beta)^{-1}\Big\{\delta^{a3}\delta^{3b}
\,+\,\delta^{a1}\delta^{1b}\Big[1-\frac{_{\sigma_\ell}}{^{\sqrt{1
+\sigma_\ell^2}}}\sin\Big(\frac{_{s(t-t_0)}}{^{\sqrt{1+\sigma_\ell^2}}}\Big)\Big]
+\,\delta^{a2}\delta^{2b}\Big[1+\frac{_{\sigma_\ell}}{^{\sqrt{1
+\sigma_\ell^2}}}\sin\Big(\frac{_{s(t-t_0)}}{^{\sqrt{1+\sigma_\ell^2}}}\Big)\cr
+4\sigma_\ell^2\sin^2\Big(\frac{_{}s}{^{2\sqrt{1+\sigma_\ell^2}}}\Big)\Big]
+\,(\delta^{a1}\delta^{2b}-\delta^{a2}\delta^{1b})
\Big[\frac{_{\sigma_\ell^2}}{^{\sqrt{1+\sigma_\ell^2}}}
\sin\Big(\frac{_{s(t-t_0)}}{^{\sqrt{1+\sigma_\ell^2}}}\Big)-2
\sigma_\ell\sin^2\Big(\frac{_{s(t-t_0)}}{^{2\sqrt{1+\sigma_\ell^2}}}
\Big)\Big]\Big\}
\qqq
that are positive matrices with constant determinant equal to 
$\,(\gamma\beta)^{-3}$.

\nappendix{D}

\noindent We give here a proof of the FDT (\ref{FDT}) around the
non-stationary equilibrium dynamics described by the Langevin equation
(\ref{equileq2}). \,On the one hand, 
the two-time dynamical correlation function is
\qq
\big\langle\,O^1(x_s)\,O^2(x_t)\,\rangle\ =\ \int \rho(x)\,O^1(x)\,P(s,x;t,y)\,
O^2(y)\,dx\,dy
\qqq
where $\,\rho(x)=Z^{-1}\ee^{-\beta\,H(x)}\,$ is the Gibbs instantaneous
PDF of the process $\,x_t\,$ satisfying the SDE (\ref{equileq2}) 
and $\,P(s,x;t,y)\,$ are the transition PDF's. 
\,Using the first of the Kolmogorov equations (\ref{Kolmo}) and integrating 
by parts, \,we infer that
\qq
\partial_s\hspace{0.025cm}\big\langle\,O^1(x_s)\,O^2(x_t)\,\rangle\ 
=\ -\int\big(L^\dagger_s
\rho\,O^1\big)(x)\,P(s,x;t,y)\,O^2(y)\,dx\,dy\,,
\qqq 
where
\qq
L_s\ =\ \big[-\beta\,d^{ij}_s(\partial_jH)\,+\,\pi^{ij}_s(\partial_jH)\,
-\,\beta^{-1}(\partial_j\pi^{ij}_s)\big]\partial_i\,
+\,\partial_id^{ij}_s\partial_j
\qqq
and $\,L_s^\dagger\rho=0$. \,Let
\qq
L^h_s\ =\ L_s\,+\,h_s\Big[\beta\,d^{ij}_s\,(\partial_jO^1)\,
-\,\pi^{ij}_s\,(\partial_jO^1)\Big]\partial_i\,,
\qqq
be the generators of the process obtained by the replacement $\,H
\to H-h_sO^1$. \,Clearly, $\,(L^h_s)^\dagger(\rho\,\ee^{\beta h_sO^1}) =0$.
\,Expanded to the first order in $\,h_s$, \,the latter equality implies
that
\qq
\beta\,L_s^\dagger(\rho\,O_1)\,
=\,-\frac{_\partial}{^{\partial h_s}}\big|_{h=0}
\,(L^h_s)^\dagger\rho\,.
\qqq
As a consequence,
\qq
&&\partial_s\hspace{0.025cm}\big\langle\,O^1(x_s)\,O^2(x_t)\,\rangle\ 
=\ \beta^{-1}\int\big(\frac{_\partial}{^{\partial h_s}}\big|_{h=0}\,
(L^h_s)^\dagger\rho\big)(x)\,P(s,x;t,y)\,O^2(y)\,dx\,dy\cr
&&=\ \beta^{-1}\int\rho(x)\,\big(\frac{_\partial}{^{\partial h_s}}\big|_{h=0}\,
L^h_s)(x)\,P(s,x;t,y)\,O^2(y)\,dx\,dy\,. 
\qqq
The right hand side is equal to 
$\ \beta^{-1}\frac{\delta}{{\delta h_s}}\big|_{h=0}\,\big\langle\,O^2(x_t)
\,\big\rangle_h\ $ so that the identity (\ref{FDT}) follows.

\end{document}